\documentclass[journal]{IEEEtran}

\usepackage{array}
\usepackage{amsmath}
\usepackage{amssymb}
\usepackage{url}
\usepackage[normalem]{ulem} 
\usepackage{mathptmx}
\usepackage{acronym}
\usepackage{booktabs}
\usepackage[compress]{cite}
\usepackage{caption}
\usepackage{subcaption}
\usepackage[switch]{lineno}
\usepackage{graphicx}
\usepackage[colorlinks,breaklinks]{hyperref} 
\hypersetup{hypertexnames=true,
  linkcolor=blue,anchorcolor=black,citecolor=blue,urlcolor=blue
}
\graphicspath{{../imglocal/}{../img/}}
\acrodef{ADC}[ADC]{Analog to Digital Converter}
\acrodef{AER}[AER]{Address Event Representation}
\acrodef{AEX}[AEX]{AER EXtension board}
\acrodef{AFM}[AFM]{Atomic Force Microscope}
\acrodef{AMDA}[AMDA]{``AER Motherboard with D/A converters''}
\acrodef{ANN}[ANN]{Attractor Neural Network}
\acrodef{API}[API]{Application Programming Interface}
\acrodef{BD}[BD]{Bundled Data}
\acrodef{CAVIAR}[CAVIAR]{Convolution AER Vision Architecture for Real-Time}
\acrodef{CCN}[CCN]{Cooperative and Competitive Network}
\acrodef{CMOL}[CMOL]{``Hybrid CMOS nanoelectronic circuits''}
\acrodef{CMOS}[CMOS]{Complementary Metal--Oxide--Semiconductor}
\acrodef{CPU}[CPU]{Central Processing Unit}
\acrodef{CV}[CV]{Coefficient of Variation}
\acrodef{DAC}[DAC]{Digital to Analog Converter}
\acrodef{DBN}[DBN]{Deep Belief Network}
\acrodef{DFA}[DFA]{Deterministic Finite Automaton}
\acrodef{DPE}[DPE]{Dynamic Parameter Estimation}
\acrodef{DPI}[DPI]{Differential Pair Integrator}
\acrodef{DRAM}[DRAM]{Dynamic Random Access Memory}
\acrodef{DR}[DR]{Dual Rail}
\acrodef{DSP}[DSP]{Digital Signal Processor}
\acrodef{DVS}[DVS]{Dynamic Vision Sensor}
\acrodef{EBL}[EBL]{Electron Beam Lithography}
\acrodef{EDVAC}[EDVAC]{Electronic Discrete Variable Automatic Computer}
\acrodef{EIN}[EIN]{Excitatory--Inhibitory Network}
\acrodef{EM}[EM]{Expectation Maximization}
\acrodef{EPSC}[EPSC]{Excitatory Post Synaptic Current}
\acrodef{EPSP}[EPSP]{Excitatory Post--Synaptic Potential}
\acrodef{FET}[FET]{Field Effect Transistor}
\acrodef{FI}[F-I]{Frequency-Current}
\acrodef{FPGA}[FPGA]{Field Programmable Gate Array}
\acrodef{FSA}[FSA]{Finite State Automaton}
\acrodef{FSM}[FSM]{Finite State Machine}
\acrodef{GPU}[GPU]{Graphical Processing Unit}
\acrodef{HH}[H\&H]{Hodgkin \& Huxley}
\acrodef{HMM}[HMM]{Hidden Markov Model}
\acrodef{HRS}[HRS]{High-Resistive State}
\acrodef{HW}[HW]{Hardware}
\acrodef{IC}[IC]{Integrated Circuit}
\acrodef{ICT}[ICT]{Information and Communication Technology}
\acrodef{IF2DWTA}[IF2DWTA]{Integrate \& Fire 2--Dimensional WTA}
\acrodef{IFSLWTA}[IFSLWTA]{Integrate \& Fire Stop Learning WTA}
\acrodef{IF}[I\&F]{Integrate-and-Fire}
\acrodef{INCF}[INCF]{International Neuroinformatics Coordinating Facility}
\acrodef{INI}[INI]{Institute of Neuroinformatics}
\acrodef{IO}[I/O]{Input/Output}
\acrodef{IPSC}[IPSC]{Inhibitory Post-Synaptic Current}
\acrodef{IPSP}[IPSP]{Inhibitory Post-Synaptic Potential}
\acrodef{ISI}[ISI]{Inter--Spike Interval}
\acrodef{JFLAP}[JFLAP]{Java - Formal Languages and Automata Package}
\acrodef{LPF}[LPF]{Low-Pass Filter}
\acrodef{LRS}[LRS]{Low-Resistive State}
\acrodef{LSM}[LSM]{Liquid State Machine}
\acrodef{LTD}[LTD]{Long Term Depression}
\acrodef{LTI}[LTI]{Linear Time-Invariant}
\acrodef{LTP}[LTP]{Long Term Potentiation}
\acrodef{LTU}[LTU]{Linear Threshold Unit}
\acrodef{MCMC}[MCMC]{Markov-Chain Monte Carlo}
\acrodef{MIM}[MIM]{Metal Insulator Metal}
\acrodef{MOSFET}[MOSFET]{Metal Oxide Semiconductor Field Effect Transistor}
\acrodef{ND}[ND]{Noise-Driven}
\acrodef{NIL}[NIL]{Nano-Imprint Lithography}
\acrodef{NMDA}[NMDA]{N-Methyl-D-Aspartate}
\acrodef{NME}[NE]{Neuromorphic Engineering}
\acrodef{PCB}[PCB]{Printed Circuit Board}
\acrodef{PSC}[PSC]{Post-Synaptic Current}
\acrodef{PSTH}[PSTH]{Peri-Stimulus Time Histogram}
\acrodef{RNN}[RNN]{Recurrent Neural Network}
\acrodef{RRAM}[ReRAM]{Resistive Random Access Memory}
\acrodef{RMSE}[RMSE]{Root Mean Squared-Error}
\acrodef{SAC}[SAC]{Selective Attention Chip}
\acrodef{SCX}[SCX]{Silicon CorteX}
\acrodef{SD}[SD]{Signal-Driven}
\acrodef{SEM}[SEM]{Spike-based Expectation Maximization}
\acrodef{SOI}[SOI]{Silicon on Insulator}
\acrodef{SRAM}[SRAM]{Static Random Access Memory}
\acrodef{STDP}[STDP]{Spike--timing--dependent Plasticity}
\acrodef{STD}[STD]{Short-Term Depression}
\acrodef{SW}[SW]{Software}
\acrodef{VHDL}[VHDL]{VHSIC Hardware Description Language}
\acrodef{VLSI}[VLSI]{Very Large Scale Integration}
\acrodef{WTA}[WTA]{Winner-Take-All}
\acrodef{divmod3}[DIVMOD3]{divisibility of a number by 3}
\acrodef{hWTA}[hWTA]{Hard Winner--Take--All}
\acrodef{sWTA}[sWTA]{soft Winner-Take-All}
\acrodef{NDFSM}[NDFSM]{Non-deterministic Finite State Machine}

\newcommand{\revised}[3]{#2}

\begin{document}



%
\title{Neuromorphic electronic circuits for building autonomous
  cognitive systems}
\author{Elisabetta~Chicca,~\IEEEmembership{Member,~IEEE,}
        Fabio~Stefanini,
        Chiara Bartolozzi,~\IEEEmembership{Member,~IEEE}
        and~Giacomo~Indiveri~\IEEEmembership{Senior Member,~IEEE}
\thanks{E. Chicca is with the Cognitive Interaction Technology
- Center of Excellence, Bielefeld University and Faculty of Technology, Bielefeld, Germany email:chicca[at]cit-ec.uni-bielefeld.de}
\thanks{C. Bartolozzi is with the iCub Facility, Istituto Italiano di Tecnologia, Genova, Italy}
\thanks{F. Stefanini and G. Indiveri are with the Institute of Neuroinformatics, University of Zurich
and ETH Zurich, Switzerland}
\thanks{Manuscript received Month DD, YEAR; revised MONTH DD, YEAR.}}

\markboth{Proceedings of the IEEE,~Vol.~x, No.~x, Month~year}%
{Chicca \MakeLowercase{\textit{et al.}}: Neuromorphic electronic circuits for building autonomous
  cognitive systems}
%

\maketitle

\begin{abstract}
  Several analog and digital brain-inspired electronic systems have
  been recently proposed as dedicated solutions for fast simulations
  of spiking neural networks. While these architectures are useful for
  exploring the computational properties of large-scale models of the
  nervous system, the challenge of building
  \revised{electronic}{low-power compact}{intelligent} physical
  artifacts that can behave \revised{intelligent}{intelligently}{} in
  the real-world and exhibit cognitive abilities still remains open.
  In this paper we propose a set of neuromorphic engineering solutions
  to address this challenge.  In particular, we review neuromorphic
  circuits for emulating neural and synaptic dynamics in real-time and
  discuss the role of biophysically realistic temporal dynamics in
  hardware neural processing architectures; we review the challenges
  of realizing spike-based plasticity mechanisms in real physical
  systems and present examples of analog
  \revised{VLSI}{electronic}{\acs{VLSI}} circuits that implement them;
  we describe the computational properties of recurrent neural
  networks and show how neuromorphic Winner-Take-All
  \revised{wta}{circuits}{recurrent network architectures} can
  implement working-memory and decision-making mechanisms. We validate
  the \revised{nm1}{neuromorphic}{} approach \revised{nm2}{proposed}{}
  with experimental results obtained from our own circuits and
  systems, and argue how the circuits and networks
  \revised{described}{presented in this work}{proposed} represent a
  useful set of components for efficiently and elegantly implementing
  neuromorphic cognition.
\end{abstract}


%
\IEEEpeerreviewmaketitle

\section{Introduction} %
\label{sec:introduction}

Machine simulation of cognitive functions has been a challenging research field
since the advent of digital computers. However, despite the large efforts and
resources dedicated to this field, humans, mammals, and many other animal
species including insects, still outperform the most powerful computers in
relatively routine functions such as sensory processing, motor control and
pattern recognition. The disparity between conventional computing technologies
and biological nervous systems is even more pronounced for tasks involving
autonomous real-time interactions with the environment, especially in presence
of noisy and uncontrolled sensory input. One important aspect is that the
computational and organizing principles followed by the nervous system are
fundamentally different from those of present day computers. Rather than using
Boolean logic, precise digital representations and clocked operations, nervous
systems carry out robust and reliable computation using hybrid analog/digital
unreliable processing elements; they emphasize distributed, event-driven,
collective, and massively parallel mechanisms and make extensive use of
adaptation, self-organization and learning.

Several approaches have been recently proposed for building custom
hardware, brain-like neural processing architectures~\cite{Jin_etal10,
  Schemmel_etal08, Silver_etal07, Wijekoon_Dudek12, Brink_etal13,
  Painkras_etal13, Pfeil_etal13, Cruz-Albrecht_etal13,
  Giulioni_etal12}. The majority of them are proposed as an
alternative electronic substrate to traditional computing
architectures for neural \emph{simulations}~\cite{Brink_etal13,
  Schemmel_etal08, Pfeil_etal13, Wijekoon_Dudek12}. These systems can
be very useful tools for neuroscience modeling, e.g., by accelerating
the simulation of complex computational neuroscience models by three
or more orders of magnitude~\cite{Pfeil_etal13, Wijekoon_Dudek12,
  Schmuker_etal14}.  However, our work focuses on an alternative
approach aimed at the realization of compact, real-time, and energy
efficient computational devices that directly \emph{emulate} the style
of computation of the brain, using the physics of Silicon to reproduce
the bio-physics of the neural tissue. This approach, on one hand,
leads to the implementation of compact and low-power behaving systems
ranging from brain-machine interfaces to autonomous robotic agents. On
the other hand, it serves as a basic research instrument for exploring
the computational properties of the neural system they emulate and
hence gain a better understanding of its operational principles. These
ideas are not new: they follow the original vision of
Mead~\cite{Mead90}, Mahowald~\cite{Mahowald92}, and
colleagues~\cite{Douglas_etal95b}. Indeed, analog \ac{CMOS} technology
has been effectively employed for the construction of simple
neuromorphic circuits reproducing basic dynamical properties of their
biological counterparts, e.g., neurons and synapses, at some level of
precision, reliability and detail. These circuits have been integrated
into \ac{VLSI} devices for building real-time sensory-motor systems
and robotic demonstrators of neural computing
architectures~\cite{Horiuchi_Koch99,Indiveri_Douglas00, Indiveri01b,
  Bartolozzi_Indiveri09b,
  Lewis_etal03,Serrano-Gotarredona_etal09}. However, these systems,
synthesized using \emph{ad-hoc} methods, could only implement very
specific sensory-motor mappings or functionalities. The challenge that
remains open is to bridge the gap from designing these types of
\emph{reactive} artificial neural modules to building complete
neuromorphic behaving systems that are endowed with cognitive
abilities.  The step from reaction to cognition in neuromorphic
systems is not an easy one, because the principles of cognition remain
to be unraveled. A formal definition of these principles and their
effective implementation in hardware is now an active domain of
research~\cite{Neftci_etal13, Eliasmith_etal12, Cassidy_etal13b,
  Cassidy_etal13}. The construction of brain-like processing systems
able to solve cognitive tasks requires sufficient theoretical grounds
for understanding the computational properties of such a system (hence
its necessary components), and effective methods to combine these
components in neuromorphic systems. During the last decade we pursued
this goal by realizing neuromorphic electronic circuits and systems
and using them as building blocks for the realization of simple
neuromorphic cognitive systems~\cite{Neftci_etal13}. Here we describe
these circuits, analyze their dynamics in comparison with other
existing solutions and present experimental results that demonstrate
their functionalities.  We describe the limitations and problems of
such circuits, and propose effective design strategies for building
larger brain-like processing systems. We conclude with a discussion on
the advantages and disadvantages of the approach we followed and with
a description of the challenges that need to be addressed in order to
progress in this domain. Specifically, in the following sections we
show how the building blocks we propose, based on dynamic synapse
circuits, hardware models of spiking neurons, and spike-based
plasticity circuits, can be integrated to form multi-chip spiking
recurrent and Winner-Take-All neural networks, which in turn have been
proposed as neural models for explaining pattern
recognition~\cite{Senn_Fusi05, Brader_etal07}, working
memory~\cite{Giulioni_etal12, Renart_etal03b}, decision
making~\cite{Deco_Rolls05, Schoner08} and state-dependent
computation~\cite{Rutishauser_Douglas09, Rigotti_etal10} in the brain.

\section{Neural dynamics in analog VLSI} %
\label{sec:dpi}

Unlike a von Neumann computing architecture, neuromorphic
architectures are composed of massively parallel arrays of simple
processing elements in which memory and computation are
co-localized. In these architectures time represents itself and so the
synapse and neuron circuits must process input data on demand, as they
arrive, and must produce their output responses in
real-time. Consequently, in order to interact with the environment and
process real-world sensory signals efficiently, neuromorphic behaving
systems must use circuits that have biologically plausible time
constants (i.e., of the order of tens of milliseconds). In this way,
they are well matched to the signals they process and are inherently
synchronized with the real-world events. This constraint is not easy
to satisfy using analog \ac{VLSI} technology. Standard analog circuit
design techniques either lead to bulky and silicon-area expensive
solutions~\cite{Rachmuth_etal11} or fail to meet this condition,
resorting to modeling neural dynamics at ``accelerated'' unrealistic
time-scales~\cite{Wijekoon_Dudek08, Schemmel_etal07}.

One elegant solution to this problem is to use current-mode design
techniques~\cite{Tomazou_Lidgey_etal90} and log-domain
\emph{subthreshold} circuits~\cite{Drakakis_etal99,
  Edwards_Cauwenberghs00, Liu_etal02b, Yu_Cauwenberghs10b,
  Mitra_etal10}. When \acp{MOSFET} are operated in the subthreshold
domain, the main mechanism of carrier transport is that of diffusion,
as it is for ions flowing through proteic channels across neuron
membranes.  As a consequence, \acp{MOSFET} have an exponential
relationship between gate-to-source voltage and drain current, and
produce currents that range from femto- to
nano-Amp\`eres. As the time constants of
  log-domain circuits are inversely proportional to their reference
  currents, in addition to being directly proportional to the circuit
  capacitance, the subthreshold domain allows the integration of
relatively small capacitors in \ac{VLSI} to implement temporal filters
that are both compact and have biologically realistic time constants,
ranging from tens to hundreds of milliseconds.

\begin{figure}[t]
  \centering
  \begin{subfigure}[h]{0.35\textwidth}
    \center{\includegraphics[height=0.55\textwidth]{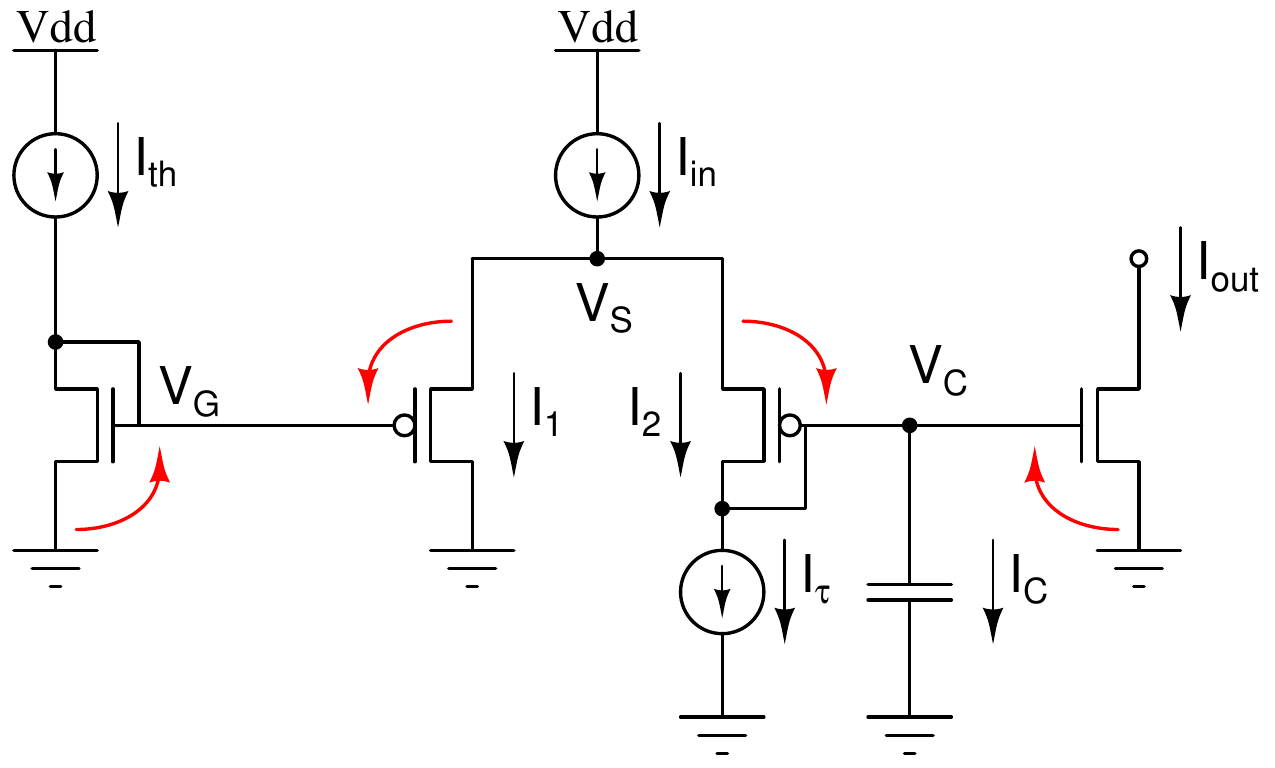}}
    \subcaption{}
    \label{fig:dpi}
  \end{subfigure} \\
    \vspace{1em}
  \begin{subfigure}[h]{0.35\textwidth}
    \center{\includegraphics[height=0.55\textwidth]{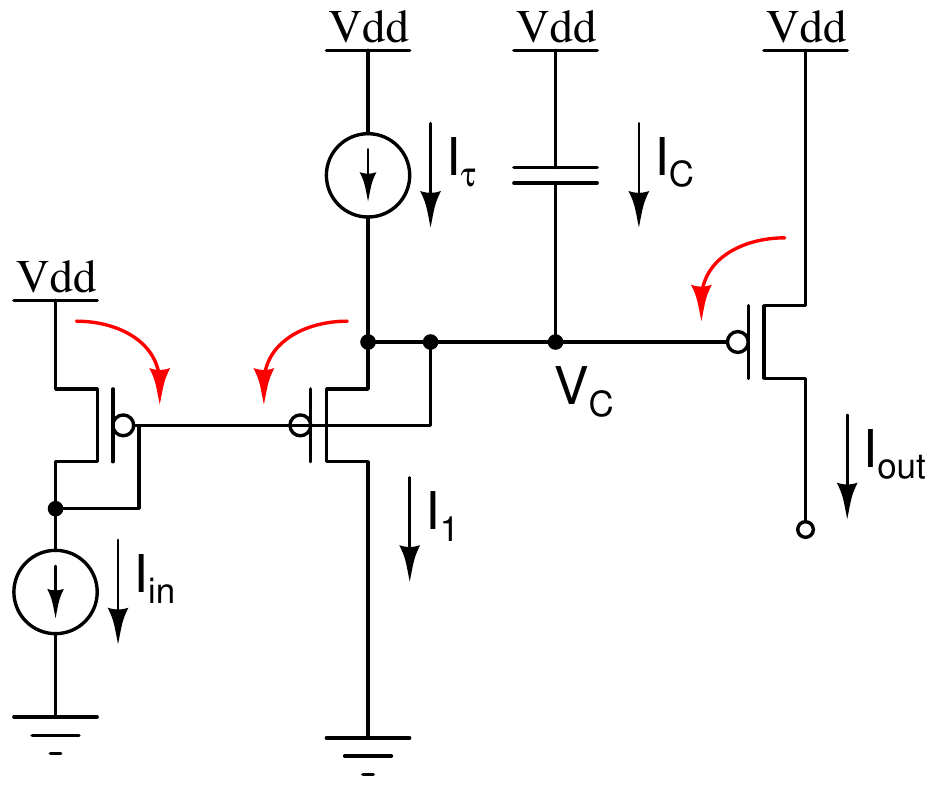}}
    \subcaption{}
    \label{fig:lpf}
  \end{subfigure} \\
    \vspace{1em}
  \begin{subfigure}[h]{0.35\textwidth}
    \center{\includegraphics[height=0.55\textwidth]{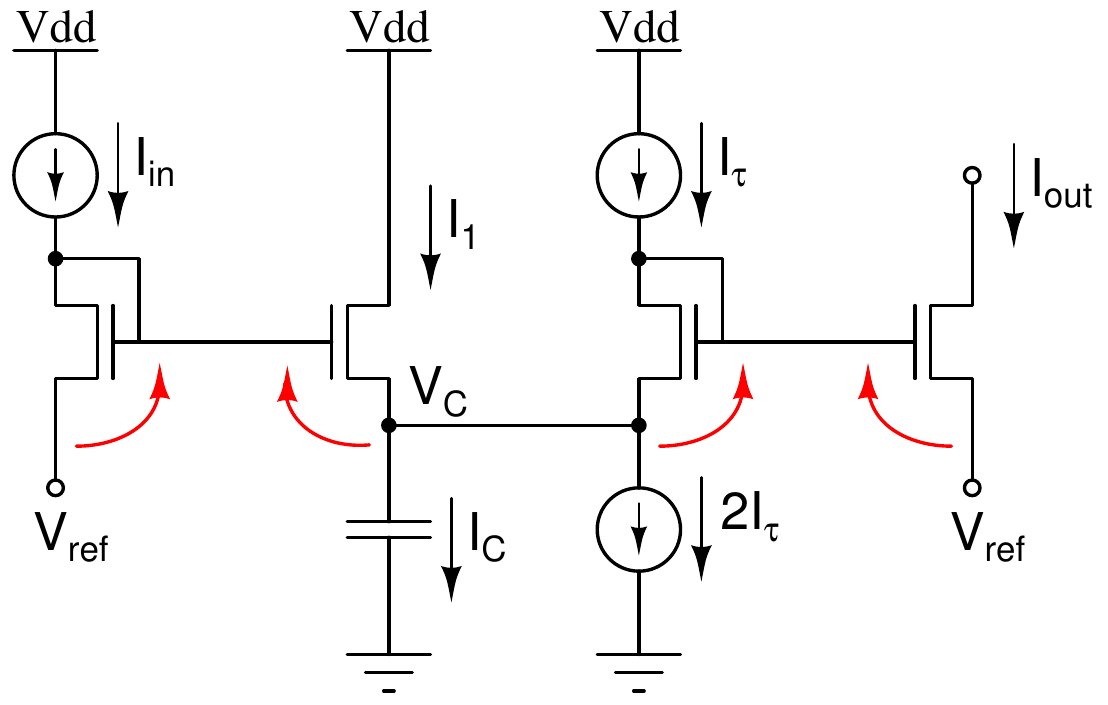}}
    \subcaption{}
    \label{fig:taucell}
  \end{subfigure}\\
  \caption{Current-mode low-pass filter circuits. Red arrows show the
    translinear loop considered for the log-domain
    analysis. (\subref{fig:dpi}) The \acl{DPI} circuit
    diagram. (\subref{fig:lpf}) The \acl{LPF} circuit diagram.
    (\subref{fig:taucell}) The ``Tau-Cell'' circuit diagram. }
  \label{fig:logdomain}
\end{figure}

Neuron conductance dynamics and synaptic transmission can be
faithfully modeled by first order differential
equations~\cite{Destexhe_etal98}, therefore subthreshold log-domain
circuits that implement first order low pass filters can faithfully
reproduce biologically plausible temporal dynamics. Several examples
of such circuits have been proposed as basic building blocks for the
implementation of silicon neurons and synapses. Among them, the
\ac{DPI}~\cite{Bartolozzi_Indiveri07, Bartolozzi_etal06}, the
log-domain \ac{LPF}~\cite{Arthur_Boahen04}, and the
``Tau-Cell''~\cite{Schaik_Jin03} circuits offer a compact and
low-power solution.  These circuits, shown in
Fig.~\ref{fig:logdomain}, can be analyzed by applying the
\emph{translinear principle}, whereby the sum of voltages in a chain
of transistors that obey an exponential current-voltage characteristic
can be expressed as multiplication of the currents flowing across
them~\cite{Gilbert96}.  For example, if we consider the \ac{DPI}
circuit of Fig.~\ref{fig:dpi}, and we assume that all transistor have
same parameters and operate in the subthreshold regime and in
saturation~\cite{Liu_etal02b}, we can derive circuit solution
analytically. Specifically, we can write:
  \begin{eqnarray}
    \label{eq:dpi}
    I_{out} = I_{0} e^{\frac{\kappa V_{C}}{U_{T}}} & \; \; &  I_{C} = C \frac{d}{dt}{V_{C}}\\
    I_{in} = I_{1} + I_{2} & \; \; & I_{2} = I_{\tau} + I_{C} \nonumber
  \end{eqnarray}
  where the term $I_0$ represents the transistor dark current, $U_{T}$
  represents the thermal voltage and $\kappa$ the subthreshold slope
  factor~\cite{Liu_etal02b}.  By
  applying the translinear principle across the loop made by the
  arrows in the circuit diagram of Fig.~\ref{fig:dpi} we can write:
$I_{th}~\cdot~I_{1}~=~I_{2}~\cdot~I_{out}$.  Then, by replacing
$I_{1}$ and expanding $I_{2}$ from eq.\,(\ref{eq:dpi}) we get:
\begin{equation}
I_{th} \cdot
(I_{in} - I_{\tau} - I_{C}) = (I_{\tau} + I_{C}) \cdot I_{out}.
\label{eq:tloop}
\end{equation}

Thanks to the properties of exponential functions, we can express
$I_{C}$ as a function of $I_{out}$:
\begin{equation}
I_{C} = C\frac{U_{T}}{\kappa I_{out}}\frac{d}{dt}I_{out}
\label{eq:ic2}
\end{equation}
Finally, by replacing $I_{C}$ from this equation and dividing everything by
$I_{\tau}$ in eq.\,(\ref{eq:tloop}), we get:

\begin{equation}
   \tau \left( 1 + \frac{I_{th}}{I_{out}} \right) \frac{d}{dt}I_{out} + I_{out} = \frac{I_{th} I_{in}}{I_{\tau}} - I_{th}
  \label{eq:nonlinear}
\end{equation}
where $\tau  \triangleq {C U_{T}}/{\kappa I_{\tau}}$.

This is a first-order \emph{non-linear} differential equation that
cannot be solved explicitly. However, in the case of sufficiently
large input currents (i.e., $I_{in} \gg I_{\tau}$) the term $-I_{th}$
on the right side of eq.~(\ref{eq:nonlinear}) can be neglected.
Furthermore, under this assumption and starting from an initial
condition $I_{out}=0$, $I_{out}$ will increase monotonically and
eventually the condition $I_{out} \gg I_{th}$ will be met.  In this
case also the term $\frac{I_{th}}{I_{out}}$ on the left side of
eq.~(\ref{eq:nonlinear}) can be neglected.  So the response of the
\ac{DPI} reduces to a first-order linear differential equation:
\begin{equation}
  \label{eq:linear}
  \boxed{\tau \frac{d}{dt}I_{out} + I_{out} = \frac{I_{th}}{I_{\tau}}I_{in}}
\end{equation}

The general solution of the other two log-domain circuits shown in
Fig.~\ref{fig:lpf} and Fig.~\ref{fig:taucell} can be derived
analytically following a similar procedure. Table~\ref{tab:syn} shows
the equations used for the derivation of all three circuits, and their
general solution.

\begin{table}
\begin{tabular}{@{}l l l@{}}
\toprule
\textbf{\acs{DPI}}&\textbf{\acs{LPF}}&\textbf{Tau-Cell}\\
\midrule
\multicolumn{3}{l}{\textbf{Circuit equations}}\\
$I_{out} = I_{0}e^{\frac{\kappa V_{C}}{U_T}}$ &
$I_{out} = I_{0}e^{\frac{\kappa (V_{dd}-V_{C})}{U_T}}$ &
$I_{out} = I_{0}e^{\frac{\kappa V_{2}-V_{ref}}{U_T}}$\\
&&\\
$I_{C} = C\frac{dV_{C}}{dt}$ &
$I_{C} = -C\frac{dV_{C}}{dt}$ &
$I_{C} = C\frac{dV_{C}}{dt}$\\
&&\\
$I_{in} = I_{1}+I_{\tau}+I_{C}$ &$I_{1}=I_{\tau}+I_{C}$ &$I_{1}=I_{\tau}+I_{C}$ \\
&&\\
$I_{C} = C\frac{U_T}{\kappa I_{out}}\frac{dI_{out}}{dt}$ &
$I_{C} = C\frac{U_T}{\kappa I_{out}}\frac{dI_{out}}{dt}$ &
$I_{C} = C\frac{U_T}{I_{out}}\frac{dI_{out}}{dt}$\\
&&\\
\midrule
\multicolumn{3}{l}{\textbf{Translinear Loop}}\\
&&\\
$I_{th} \cdot I_{1} = (I_{\tau}+I_{C}) \cdot I_{out}$ &
$I_{in} \cdot I_{0} = I_{1} \cdot I_{out}$ &
$I_{in}  \cdot I_{\tau} =  I_{1} \cdot I_{out}$\\
&&\\
\midrule
\multicolumn{3}{l}{\textbf{Solution}}\\
&&\\
$\displaystyle\tau \frac{dI_{out}}{dt} + I_{out} = \frac{I_{th}}{I_{\tau}} I_{in}\;\;\;$ &
$\displaystyle\tau \frac{dI_{out}}{dt} + I_{out} = \frac{I_{0}}{I_{\tau}} I_{in}\;\;\;$ &
$\displaystyle\tau \frac{dI_{out}}{dt} + I_{out} = I_{in}$\\
&&\\
$\tau = \frac{CU_T}{\kappa I_{\tau}}$ &
$\tau = \frac{CU_T}{\kappa I_{\tau}}$ &
$\tau = \frac{CU_T}{I_{\tau}}$\\
\bottomrule
\end{tabular}
\caption{Characteristic equations of the \acs{DPI}, \acs{LPF}, and Tau-Cell log-domain  filters.}\label{tab:syn}
\end{table}
The \ac{LPF} circuit of Fig.~\ref{fig:logdomain} is the one that has
the least number of components. However it is not the most compact,
because to apply the translinear principle correctly, it is necessary
to use a p-type \ac{FET} with its bulk connected to its source node
(see p-\ac{FET} with $I_{1}$ current flowing through it in
Fig.~\ref{fig:lpf}). This requires an isolated well in the circuit
layout, which 
leads to larger area usage, and makes the overall size of the circuit
comparable to the size of the other two solutions. Furthermore, the
requirement of an isolated well for the p-\ac{FET} does not allow the
design of the complementary version of the circuit in standard CMOS
processes (e.g., to have negative currents).  The Tau-Cell circuit
does not have this problem, but it requires precise matching of the
two current sources producing $I_{\tau}$ and $-2I_{\tau}$, which can
also lead to large area usage at the layout level.  The \ac{DPI} can
implement in a compact way both positive and negative currents (e.g.,
by using the complementary version of the schematic of
Fig.~\ref{fig:dpi}). An other advantage of the \ac{DPI}, with respect
to the other two solutions, is the availability of the additional
control parameter $I_{th}$ that can be used to change the gain of the
filter.

The \ac{LPF} circuit has been used to model both synaptic excitation
and shunting inhibition~\cite{Arthur_Boahen07}. The Tau-Cell has been
used to implement log-domain implementation of
Mihalas-Niebur~\cite{Schaik_etal10} and
Izhikevich~\cite{Schaik_etal10b} neuron models, and the \ac{DPI} has
been used to implement both synapse and neuron
models~\cite{Bartolozzi_Indiveri07,Indiveri_etal10}. In the next
sections we will show examples of neurons and synapses that exploit
the properties of the \ac{DPI} to implement the relevant dynamics.

\section{Silicon neurons} %
\label{sec:silicon-neurons}

Several \ac{VLSI} implementations of conductance-based models of
neurons have been proposed in the past~\cite{Mahowald_Douglas91,
  Dupeyron_etal96, Alvado_etal04, Simoni_etal04,Yu_Cauwenberghs10}.
Given their complexity, these circuits require significant silicon
real-estate and a large number of bias voltages or currents to
configure the circuit properties. Simplified \ac{IF} models typically
require far less transistors and parameters but often fail at
reproducing the rich repertoire of behaviors of more complex
ones~\cite{Izhikevich03, Brette_Gerstner05}.

A recently proposed class of \emph{generalized} \ac{IF}
models however has been shown to capture many of the properties
of biological neurons, while requiring less and simpler differential
equations compared to more elaborate conductance-based models, such as
the \ac{HH} one~\cite{Brette_Gerstner05, Mihalas_Niebur09}. Their
computational simplicity and compactness make them valuable options
for VLSI implementations~\cite{Wijekoon_Dudek08, Folowosele_etal09,
  Livi_Indiveri09, Schaik_etal10, Schaik_etal10b}.

We describe here a generalized \ac{IF} neuron circuit originally
presented in~\cite{Livi_Indiveri09}, which makes use of the \ac{DPI}
circuit described in the previous Section and which represents an
excellent compromise between circuit complexity and computational power:
the circuit is compact, both in terms of transistor count and layout
size; it is low-power; it has biologically realistic time constants;
and it implements refractory period and \emph{spike-frequency
adaptation}, which are key ingredients for producing resonances and
oscillatory behaviors often emphasized in more complex
models~\cite{Izhikevich03, Mihalas_Niebur09}.

\begin{figure}
  \centering
  \includegraphics[width=0.475\textwidth]{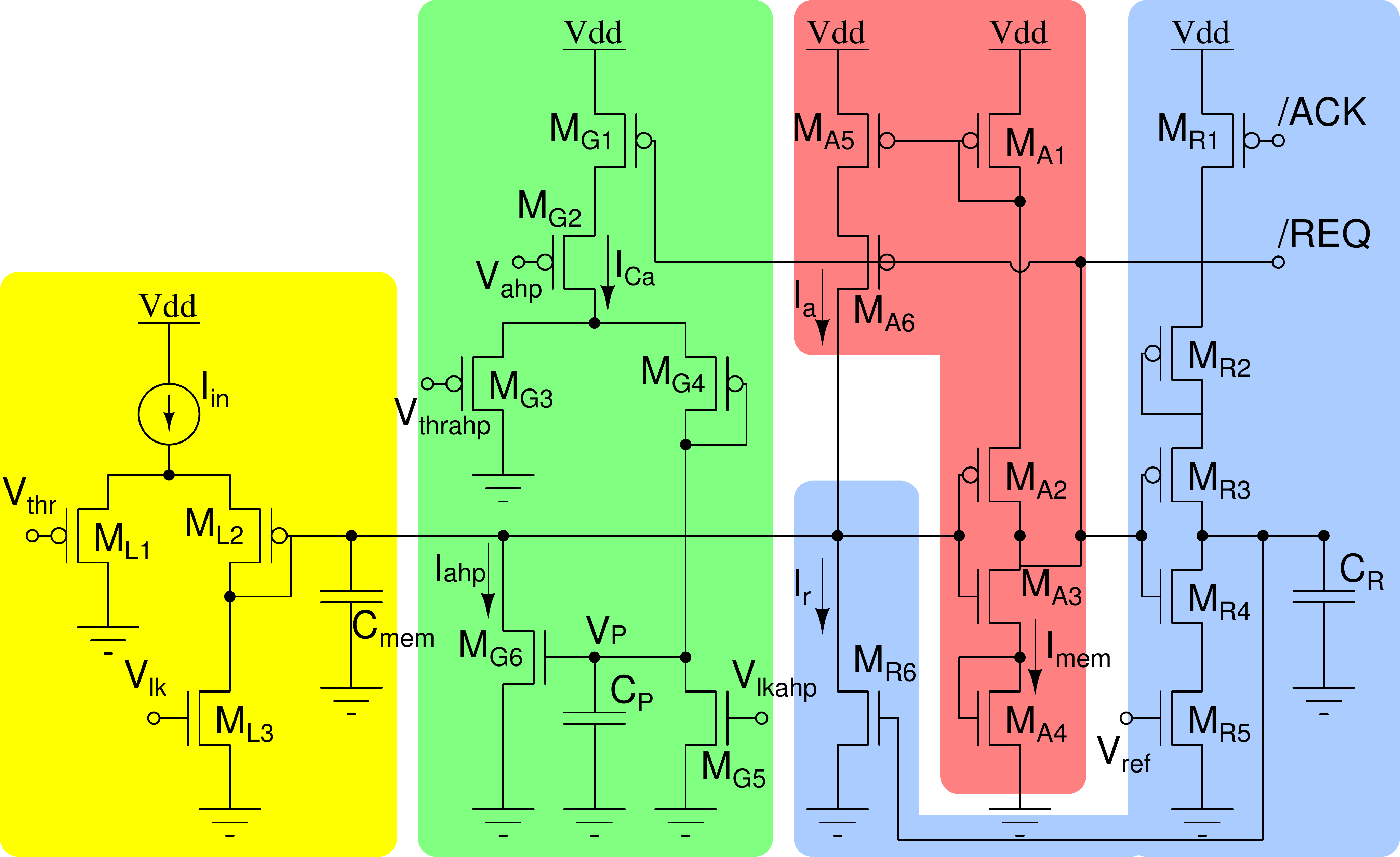}
  \caption{Adaptive exponential \ac{IF} neuron circuit schematic. The
    input \ac{DPI} circuit (${\rm M_{L1-3}}$) models the neuron's leak
    conductance. A spike event generation amplifier (${\rm M_{A1-6}}$)
    implements current-based positive feedback (modeling both sodium
    activation and inactivation conductances) and produces
    address-events at extremely low-power operation. The reset block
    (${\rm M_{R1-6}}$) resets the neuron and keeps it in a resting state
    for a refractory period, set by the ${\rm V_{ref}}$ bias
    voltage. An additional low-pass filter (${\rm M_{G1-6}}$)
    integrates the spikes and produces a slow after hyper-polarizing
    current ${\rm I_{ahp}}$ responsible for spike-frequency adaptation.}
\label{fig:dpineur}
\end{figure}

The circuit schematic is shown in Fig.~\ref{fig:dpineur}. It comprises
an input \ac{DPI} circuit used as a low-pass filter (${\rm
  M_{L1-3}}$), a spike-event generating amplifier with current-based
positive feedback (${\rm M_{A1-6}}$), a spike reset circuit with
refractory period functionality (${\rm M_{R1-6}}$) and a
spike-frequency adaptation mechanism implemented by an additional
\ac{DPI} low-pass filter (${\rm M_{G1-6}}$). The \ac{DPI} block ${\rm
  M_{L1-3}}$ models the neuron's leak conductance; it produces
exponential sub-threshold dynamics in response to constant input
currents. The neuron's membrane capacitance is represented by the
capacitor ${\rm C_{mem}}$ while Sodium channel activation and
inactivation dynamics are modeled by the positive-feedback circuits in
the spike-generation amplifier ${\rm M_{A1-6}}$. The reset ${\rm
  M_{R1-6}}$ block models the Potassium conductance and refractory
period functionality. The spike-frequency adaptation block ${\rm
  M_{G1-6}}$ models the neuron's Calcium conductance that produces the
after-hyper-polarizing current ${\rm I_{ahp}}$, which is proportional
to the neuron's mean firing rate.

By applying the current-mode analysis of
Section~\ref{sec:dpi} to both the input and the spike-frequency
adaptation \ac{DPI} circuits we derive the complete equation that
describes the neuron's subthreshold behavior:

\begin{align}
\label{eq:iboth}
  \left ( 1 + \frac{I_{th}}{I_{mem}} \right ) \tau\frac{d}{dt}I_{mem}   + I_{mem} \left ( 1+
  \frac{I_{ahp}}{I_{\tau}} \right ) & = I_{mem_\infty} +  f \left ( I_{mem} \right ) \nonumber \\
  \tau_{ahp}\frac{d}{dt}I_{ahp}  + I_{ahp} & =  I_{{ahp}_\infty} u(t)
\end{align}
where ${I_{mem}}$ is the sub-threshold current that represents
the real neuron's membrane potential variable,
${I_{ahp}}$ is the slow variable responsible for the
spike-frequency adaptation mechanisms, and $u(t)$ is a step function
that is unity for the period in which the neuron spikes and null in
other periods. The term $f{(I_{mem})}$ is a function that
depends on both the membrane potential variable ${I_{mem}}$ and
the positive-feedback current ${I_{a}}$ of
Fig.~\ref{fig:dpineur}:
\begin{equation}
f(I_{mem}) = \frac{I_{a}}{I_{\tau}}(I_{mem} + I_{th})
\label{eq:ifb}
\end{equation}
In~\cite{Indiveri_etal10} the authors measured ${I_{mem}}$
experimentally and showed how $f{(I_{mem})}$ could be fitted with an
exponential function of ${I_{mem}}$. The other parameters of
eq.\,(\ref{eq:iboth}) are defined as:
\begin{align*}
\tau & \triangleq \frac{C_{mem}U_T}{\kappa I_{\tau}}, & \;\;\;
\tau_{ahp} & \triangleq \frac{C_{p}U_T}{\kappa I_{\tau_{ahp}}} \\
I_{\tau} & \triangleq I_0e^{\frac{\kappa}{U_T} V_{lk}}, & \;\;\;
I_{\tau_{ahp}} & \triangleq I_0e^{\frac{\kappa}{U_T} V_{lkahp}} \\
I_{mem_\infty} & \triangleq \frac{I_{th}}{I_{\tau}} (I_{in}-I_{ahp}-I_{\tau}), & \;\;\;
I_{{ahp}_{\infty}} & \triangleq  \frac{I_{th_{ahp}}}{I_{\tau_{ahp}}}  I_{Ca}
\end{align*}
where ${I_{th}}$ and $I_{\tau_{ahp}}$ represent currents through
n-type \acp{MOSFET} not present in Fig.~\ref{fig:dpineur}, and defined as
$I_{th}\triangleq~I_0e^{\frac{\kappa}{U_T} V_{thr}}$, and
$I_{th_{ahp}}\triangleq~I_0e^{\frac{\kappa}{U_T}V_{thrahp}}$
respectively.

In addition to emulating Calcium-dependent after-hyper\-pola\-rization
Potassium currents observed in real neurons~\cite{Connors_etal82}, the
spike-frequency adaptation block ${\rm M_{G1-6}}$ reduces power
consumption and bandwidth usage in networks of these neurons. For
values of ${I_{in}} \gg {I_{\tau}}$ we can make the same simplifying
assumptions made in Section~\ref{sec:dpi}. Under these assumptions,
and ignoring the adaptation current ${I_{ahp}}$, eq.~(\ref{eq:iboth})
reduces to:
\begin{equation}
\label{eq:simplifieddpineur}
\tau \frac{d}{dt}I_{mem}   + I_{mem}  = \frac{I_{th}}{I_{\tau}}I_{in} + f{(I_{mem})}
\end{equation}
where $f{(I_{mem})} \approx \frac{I_{a}}{I_{\tau}} I_{mem}$.

So under these conditions, the circuit of Fig.~\ref{fig:dpineur}
implements a \emph{generalized \ac{IF}} neuron
model~\cite{Jolivet_etal04}, which has been shown to be extremely
versatile and capable of faithfully reproducing the action potentials
measured from real cortical
neurons~\cite{Badel_etal08, Naud_etal09}. Indeed, by changing the
biases that control the neuron's time-constants, refractory period,
and spike frequency adaptation dynamics this circuit can produce a
wide range of spiking behaviors ranging from regular spiking to
bursting (see Section~\ref{sec:experimental-results}).


While this circuit can express dynamics with time constants of
hundreds of milliseconds, it is also compatible with fast asynchronous
digital circuits (e.g., $<100$\,nanosecond pulse widths), which are
required to build large spiking neural network architectures (see the
\textsf{/REQ} and \textsf{/ACK} signals of Fig.~\ref{fig:dpineur} and
Section~\ref{sec:networks}). This allows us to integrate multiple
neuron circuits in event-based \ac{VLSI} devices and construct large
distributed re-configurable neural networks.

\section{Silicon synapses}
\label{sec:silicon-synapses}

Synapses are fundamental elements for computation and information
transfer in both real and artificial neural systems, and play a
crucial role in neural coding and learning. While modeling the
non-linear properties and the dynamics of large ensembles of synapses
can be extremely onerous for \ac{SW} simulations (e.g., in terms of
computational power and simulation time), dedicated neuromorphic
\ac{HW} can faithfully reproduce synaptic dynamics in real-time using
massively parallel arrays of pulse (spike)
integrators. In this case, the bottleneck is not
  in the complexity of the spiking processes being modeled, but in the
  number of spikes being received and transmitted (see
  Section~\ref{sec:networks} for more details).

\begin{figure}
  \centering
  \includegraphics[width=0.475\textwidth]{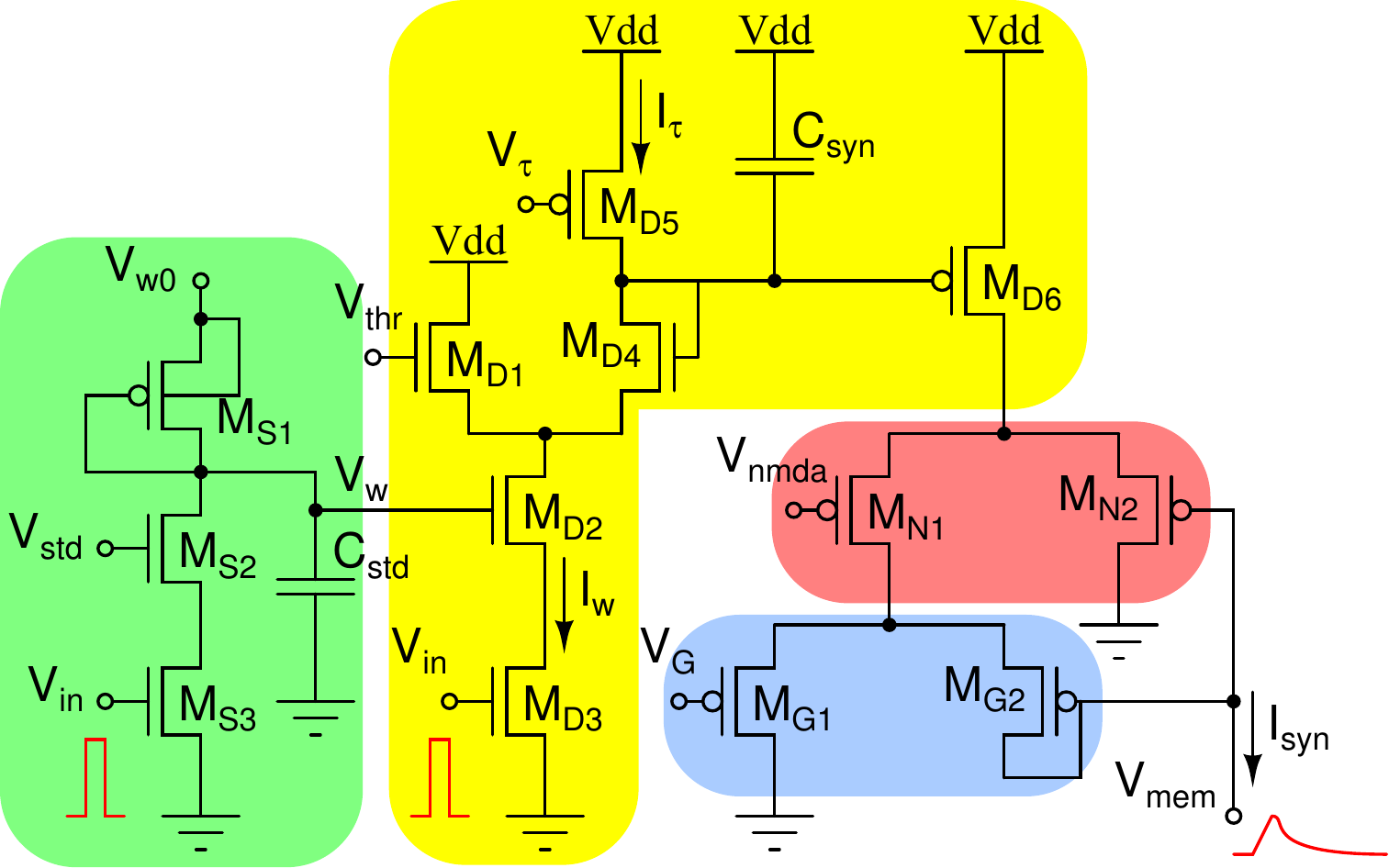}
  \caption{Complete \ac{DPI} synapse circuit, including short term
    plasticity, NMDA voltage gating, and conductance-based functional
    blocks. The short-term depression block is implemented by
    \acp{MOSFET} ${\rm M_{S1-3}}$; the basic \ac{DPI} dynamics are
    implemented by the block ${\rm M_{D1-6}}$; the \acs{NMDA} voltage
    gated channels are implemented by ${\rm M_{N1-2}}$, and conductance
    based voltage dependence is achieved with ${\rm M_{G1-2}}$.}
  \label{fig:dpi-tot}
\end{figure}

An example of a full excitatory synapse circuit is shown in
Fig.~\ref{fig:dpi-tot}. This circuit, based on the \ac{DPI} circuit
described in Section~\ref{sec:dpi}, produces  biologically realistic
\acp{EPSC}, and can express short term plasticity, \ac{NMDA} voltage
gating, and conductance-based behaviors. The input spike (the voltage pulse
$V_{in}$) is applied to both ${\rm M_{D3}}$ and ${\rm M_{S3}}$. The
output current $I_{syn}$, sourced from ${\rm M_{D6}}$ and through
${\rm M_{G2}}$, rises and decays exponentially with time.  The temporal
dynamics are implemented by the \ac{DPI} block ${\rm M_{D1-6}}$.  The
circuit time constant is set by $V_{\tau}$ while the synaptic
efficacy, which determines the \ac{EPSC} amplitude, depends on both
$V_{w0}$ and $V_{thr}$~\cite{Bartolozzi_Indiveri07}.

\subsection{Short term depression and short-term facilitation}
\label{sec:dpistp}

Short term plasticity mechanisms can be extremely effective tools for
processing temporal signals and decoding temporal
information~\cite{Buonomano00,Zucker_Regehr02}. Several circuit
solutions have been proposed to implement these types of dynamics,
using different types of devices and following a wide range of design
techniques~\cite{Rasche_Hahnloser01, Boegerhausen_etal03, Bill_etal10,
  Noack_etal11, Ohno_etal11, Dowrick_etal12}. These short-term dynamic
mechanisms are subdivided into \emph{short-term depression} and
\emph{short-term facilitation}.  The circuit block ${\rm M_{S1-3}}$ is
responsible for implementing short-term depression: with every voltage
pulse $V_{in}$ the synaptic weight voltage $V_{w}$ decreases, at a
rate set by $V_{std}$. When no spikes are being received, the $V_{w}$
``recovers'' toward the resting state set by
$V_{w0}$. In~\cite{Boegerhausen_etal03} the authors demonstrate that
this sub-circuit is functionally equivalent to the one described in
theoretical models, and often used in computational neuroscience
simulations~\cite{Abbott_etal97, Tsodyks_Markram97}.  In addition to
\emph{short-term depression}, this \ac{DPI} synapse is capable also of
\emph{short-term facilitation}: if the bias $V_{thr}$ of ${\rm
  M_{D1}}$ is set so that $I_{th} \gg I_{syn}$ at the onset of the
stimulation (i.e., during the first spikes), the circuit equation,
derived from eq.~(\ref{eq:nonlinear}) in the analysis of
Section~\ref{sec:dpi} reduces to:

\begin{equation}
  \tau \frac{d}{dt}I_{syn} + \frac{I_{syn}^2}{I_{th}} - I_{syn} ( \frac{I_{w}}{I_{\tau}} +1) = 0
  \label{eq:stf-full}
\end{equation}
 which can be further simplified to:
\begin{equation}
  \tau \frac{d}{dt}I_{syn} = I_{syn} ( \frac{I_{w}}{I_{\tau}} +1)
  \label{eq:stf}
\end{equation}

In other words, the change in circuit response increases with every
spike, by an amount greater than one, for as long as the condition
$I_{syn} \ll I_{th}$ is satisfied. As $I_{syn}$ increases this
condition starts to fail, and eventually the opposite condition
($I_{syn} \gg I_{th}$) is reached. This is the condition for
linearity, under which the circuit starts to behave as a first order
low-pass filter, as described in Section~\ref{sec:dpi}.

\subsection{NMDA voltage gating and conductance behavior}
\label{sec:dpinmda}
The output differential pairs of Fig.~\ref{fig:dpi-tot}
(${\rm M_{N1-2}}$ and ${\rm M_{G1-2}}$) are responsible for implementing
\ac{NMDA} voltage gated channels and conductance-based behavior
respectively. The response properties of these circuits have been
thoroughly characterized in~\cite{Bartolozzi_Indiveri07}.

\subsection{Homeostatic plasticity: synaptic scaling}
\label{sec:dpiscaling}
Synaptic scaling is a stabilizing homeostatic mechanism used
by biological neural systems to keep the network's activity within proper
operating bounds. It operates by globally scaling the synaptic weights of all
the synapses afferent to a neuron, for maintaining the neuron's firing rate
within a functional range, in face of chronic changes of their activity level,
while preserving the relative differences between individual
synapses~\cite{Turrigiano_etal98}. In \ac{VLSI}, synaptic scaling is an
appealing mechanism that can be used to compensate for undesired behaviors that
can arise for example because of temperature drifts or sudden changes in the
system input activity levels. Thanks to its independent controls on synaptic
efficacy set by $V_{w}$ and $V_{thr}$, the \ac{DPI} synapse of
Fig.~\ref{fig:dpi-tot} is compatible with both conventional spike-based
learning rules, and homeostatic synaptic scaling mechanisms. Specifically,
while learning circuits can be designed to locally change the synaptic weight
by acting on the $V_{w}$ of each individual synapse (e.g., see
Section~\ref{sec:lcircuits}), it is possible to implement adaptive circuits
that act on the $V_{thr}$ of all the synapses connected to a given neuron to
keep its firing rate within desired control boundaries. This strategy has been
recently demonstrated in~\cite{Bartolozzi_Indiveri09}.

\section{Synaptic plasticity: spike-based learning circuits} %
\label{sec:lcircuits}

One of the key properties of biological synapses is their ability to
exhibit different forms of \emph{plasticity}.  Plasticity mechanisms
produce long-term changes in the synaptic strength of individual
synapses in order to form memories and learn about the statistics of
the input stimuli. Plasticity mechanisms that induce changes that
increase the synaptic weights are denoted as \ac{LTP} mechanisms, and
those that induce changes that decrease synaptic weights are denoted
as \ac{LTD} mechanisms~\cite{Abbott_Nelson00}.

In neuromorphic \ac{VLSI} chips, implementations of long-term
plasticity mechanisms allow us to implement learning algorithms and
set synaptic weights automatically, without requiring dedicated
external read and write access to each individual synapse.

As opposed to the case of theory, or software simulation, the
realization of synapses in hardware imposes a set of important
physical constraints. For example synaptic weights can only have
bounded values, and with a limited (and typically small)
precision. These constraints have dramatic effects on the memory
capacity of the neural network that uses such
synapses~\cite{Amit_Fusi92, Fusi_Abbott07}. So when developing
computational models of biological synapses that will be mapped onto
neuromorphic hardware, it is important to develop plasticity
mechanisms that work with limited resolution and bounded synaptic
weights~\cite{Senn_Fusi05}.  Another important constraint that should
be taken into account when developing hardware learning systems that
are expected to operate continuously (as is the case for real-time
behaving systems) is related to the \emph{blackout
  effect}~\cite{Amit92}. Classical Hopfield networks are affected by
this effect: in Hopfield networks the memory capacity is limited, and
is related to the number of synapses available. Learning new patterns
uses memory resources and if the number of stored patterns reaches a
critical value the storage of even one single new pattern destroys the
whole memory because none of the old patterns can be recalled.
Unfortunately, this catastrophic condition is unavoidable in most
practical scenarios, since continuous, uninterrupted learning will
always lead to the blackout effect. However, it is possible to avoid
this effect, by building networks that can progressively \emph{forget}
old memories to make room for new ones, thus exhibiting the
\emph{palimpsest property}~\cite{Nadal_etal86}.  It has been
demonstrated that the optimal strategy for implementing this
palimpsest property, while maintaining a high storage capacity, is to
use synapses that have a discrete number of stable states and that
exhibit \emph{stochastic transitions} between
states~\cite{Amit_Fusi94}.  Specifically, it was demonstrated that by
modifying only a random subset of the network synapses with a small
probability, memory lifetimes increase by a factor inversely
proportional to the probability of synaptic
modification~\cite{Fusi02}. In addition, the probability of synaptic
transitions can be used as a free parameter to set the trade-off
between the speed of learning against the memory capacity.

These types of plastic synapse circuits can be implemented in a very
compact way by reducing to the minimum the resolution of the synaptic
weight (i.e., just two stable states) and using variability in the
input spike trains as the source of stochasticity for the transition
of the synaptic weights (e.g., from an \ac{LTD} to an \ac{LTP} stable
state).  The low resolution in the synaptic weights can be compensated
by redundancy (i.e., using large numbers of synapses) and the
variability in the input spike trains can be obtained by encoding
signals with the mean rates of Poisson distributed
spike-trains~\cite{Fusi_etal00, Chicca_Fusi01, Mitra_etal09}.  An
important advantage of delegating the onus of generating the
stochasticity to the input spiking activity is that no additional
circuitry is needed for the stochastic state
transitions~\cite{Seo_etal11}. Furthermore, since the spiking activity
controls the speed of learning, the network can easily switch between
a slow-learning regime (i.e., to learn pattern of mean firing rates
with uncorrelated stimuli) to a fast learning one (i.e., to learn
highly correlated patterns) without changing its internal
parameters~\cite{Chicca_Fusi01, Sheik_etal12}.

In addition to allowing compact circuit designs, these types of
plastic synapse circuits do not require precisely matched analog
devices. As the dominant source of variability lies in the (typically
Poisson distributed) input spikes driving the learning, additional
sources of variability, for example induced by device mismatch, do not
affect the main outcome of the learning process. As a consequence,
analog VLSI designers do not have to allocate precious Silicon
real-estate resources to minimize device mismatch effects in these
circuits.

\begin{figure}
  \centering
  \begin{subfigure}[b]{0.45\textwidth}
    \centering
    \includegraphics[width=.7\textwidth]{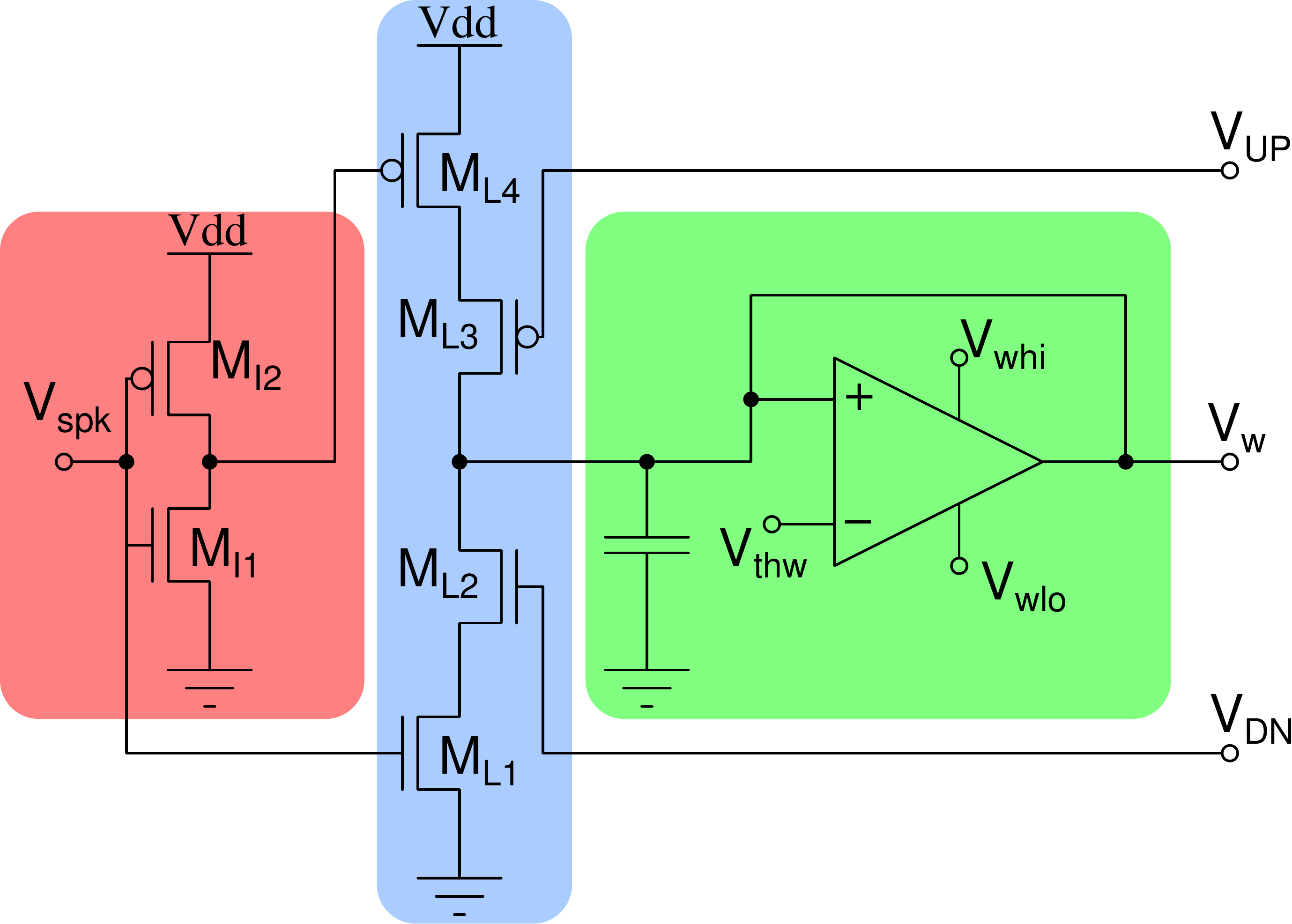} 
    \subcaption{}
    \label{fig:lsyn}
  \end{subfigure}\\
    \vspace{1em}
  \begin{subfigure}[b]{0.475\textwidth}
    \includegraphics[width=\textwidth]{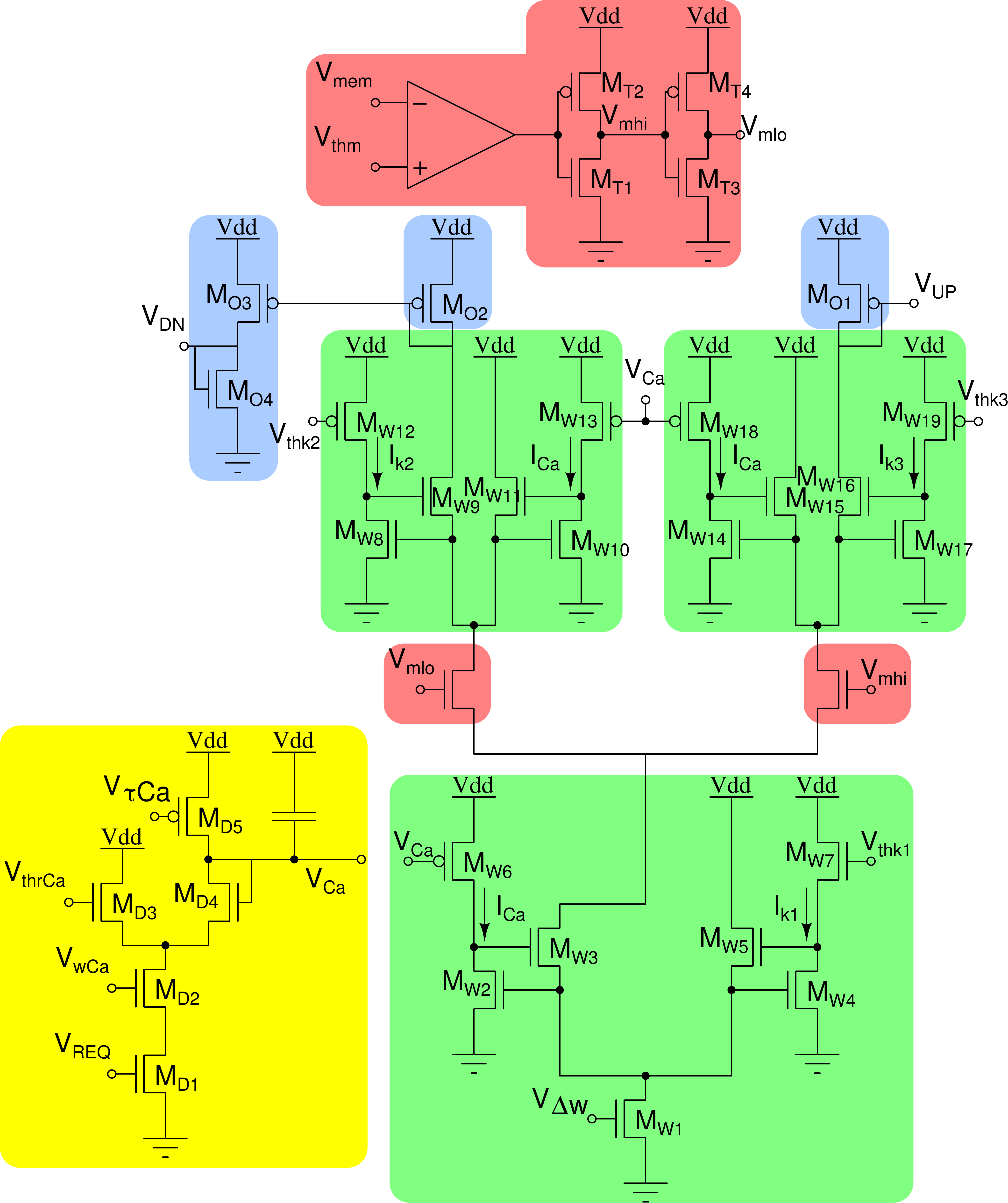}
    \subcaption{}
    \label{fig:lsoma}
  \end{subfigure}
  \caption{Spike-based learning circuits. (\subref{fig:lsyn})
    Pre-synaptic weight-update module (present at each
    synapse). (\subref{fig:lsoma}) Post-synaptic learning
    control circuits (present at the soma).}
  \label{fig:learn}
\end{figure}

An example of a circuit that implements a weight update mechanism
compatible with this stochastic learning rule, is shown in
Fig.~\ref{fig:lsyn}. The circuit comprises three main blocks: an input
stage ${\rm M_{I1-2}}$, a spike-triggered weight update block ${\rm
  M_{L1-4}}$, and a bi-stability weight storage/refresh block (see
transconductance amplifier in Fig.~\ref{fig:lsyn}). The input stage
receives spikes from pre-synaptic neurons and triggers increases or
decreases in weights, depending on the two signals $V_{UP}$ and
$V_{DN}$ generated downstream by the post-synaptic neuron. The
bi-stability weight refresh circuit is a positive-feedback amplifier
with very small ``slew-rate'' that compares the weight voltage $V_{w}$
to a set threshold $V_{thw}$ and slowly drives it toward one of the
two rails $V_{whi}$ or $V_{wlo}$, depending on whether $V_{w}>V_{thw}$
or $V_{w}<V_{thw}$ respectively. This bi-stable drive is continuous
and its effect is superimposed to the one from the spike-triggered
weight update circuit. The analog, bi-stable, synaptic weight voltage
$V_{w}$ is then used to set the amplitude of the \ac{EPSC} generated
by the synapse integrator circuit (e.g., the circuit shown in
Fig.~\ref{fig:dpi-tot}). Note that while the weight voltage $V_{w}$ is
linearly driven by the bi-stability circuit, its effect on the
\ac{EPSC} produced by the connected \ac{DPI} synapse is
exponential. This non-linearity can in principle affect adversely the
dynamics of learning and is more relevant at small scales (tens of
synapses) since the contribute of each synapse is important. However
the non-linearity has a negligible effect in practice because in the
slow-learning regime only a small subset of a much larger number of
synapses is involved in the learning process, each one participating
with a small contribution. The circuit presented here can be easily
modified to better reproduce the linear dynamics of the theoretical
model by decoupling the synaptic weight from the internal variable, as
in~\cite{Giulioni_etal09}.

The two signals $V_{UP}$ and $V_{DN}$ of Fig.~\ref{fig:lsyn} that
determine whether to increase or decrease the synaptic weight are
shared globally among all synapses afferent to a neuron. The circuits
that control these signals can be triggered by the neuron's
post-synaptic spike, to implement standard \ac{STDP} learning
rules~\cite{Abbott_Nelson00}. In general, \ac{STDP} mechanisms that
update the synaptic weight values based on the relative timing of pre-
and post-synaptic spikes can be implemented very effectively in
analog~\cite{Indiveri_etal06, Bofill-i-Petit_Murray04, Fusi_etal00,
  Hafliger_etal97, Azghadi_etal13} or mixed analog-digital \ac{VLSI}
technology~\cite{Arthur_Boahen06}.  However, while standard \ac{STDP}
mechanisms can be effective in learning to classify spatio-temporal
spike patterns~\cite{Gutig_Sompolinsky06, Arthur_Boahen06}, these
algorithms and circuits are not suitable for both encoding information
represented in a spike correlation code and a means rate code without
spike correlations~\cite{Senn02,Lisman_Spruston05}.  For this reason,
we focus on more elaborate plasticity mechanisms that not only depend
on the timing of the pre-synaptic spikes but also on other state
variables present at the post-synaptic terminal, such as the neuron
membrane potential or its Calcium concentration.  An example of such
type of learning rule is the one proposed in~\cite{Brader_etal07},
which has been shown to be able to classify patterns of mean firing
rates, to capture the rich phenomenology observed in
neurophysiological experiments on synaptic plasticity, and to
reproduce the classical \ac{STDP} phenomenology both in
hardware~\cite{Mitra_etal09, Giulioni_etal09, Giulioni_etal12} and in
software simulations~\cite{Brader_etal07, Beyeler_etal13}.  This rule
can be used to implement unsupervised and supervised learning
protocols, and to train neurons to act as perceptrons or binary
classifiers~\cite{Senn_Fusi05}. Typically, input patterns are encoded
as sets of spike trains that stimulate the neuron's input synapses
with different mean frequencies, while the neuron's output firing rate
represents the binary classifier output.

Examples of circuits that implement such a learning rule are shown in
Fig.~\ref{fig:lsoma}.
The spikes produced by the post-synaptic neuron are integrated by the
\ac{DPI} circuit ${\rm M_{D1-5}}$ to produce a voltage $V_{Ca}$ which
represents a post-synaptic Calcium concentration and is a measure of
the recent spiking activity of the neuron. The three current-mode
winner-take-all circuits~\cite{Lazzaro_etal89} ${\rm M_{W1-19}}$
compare $V_{Ca}$ to three different thresholds $V_{thk1}$, $V_{thk2}$,
and $V_{thk3}$.  In parallel, the neuron's membrane potential
$V_{mem}$ is compared to a fixed threshold $V_{thm}$ by a voltage
comparator. The outcomes of these comparisons set $V_{UP}$ and
$V_{DN}$ such that, whenever a pre-synaptic spike $V_{spk}$ reaches
the synapse weigh-update block of Fig.~\ref{fig:lsyn}:
\begin{align*}
  \begin{cases}
    V_w = V_w + \Delta w & {\rm if \;\;} V_{mem}>V_{mth} {\rm \;\; and \;\;} V_{thk1}<V_{Ca}<V_{thk3}\\
    V_w = V_w - \Delta w & {\rm if \;\;} V_{mem}<V_{mth} {\rm \;\; and \;\;} V_{thk1}<V_{Ca}<V_{thk2}
  \end{cases}
\end{align*}
where $\Delta w$ is a factor that depends on $V_{\Delta w}$ of
Fig.~\ref{fig:lsoma}, and is gated by the eligibility traces $V_{UP}$
or $V_{DN}$. If none of the conditions above are met, $\Delta w$ is
set to zero by setting $V_{UP}=V_{dd}$, and $V_{DN}=0$.

The conditions on $V_{Ca}$ implement a ``stop-learning'' mechanism
that greatly improves the generalization performance of the system by
preventing over-fitting when the input pattern has already been
learned~\cite{Senn_Fusi05,Brader_etal07}. For example, when the
pattern stored in the synaptic weights and the pattern provided in
input are highly correlated, the post-synaptic neuron will fire with a
high rate and $V_{Ca}$ will rise such that $V_{Ca}>V_{thk3}$, and no
more synapses will be modified.

In~\cite{Mitra_etal09,Giulioni_etal09} the authors show how such types
of circuits can be used to carry out classification tasks with a
supervised learning protocol, and characterize the performance of
these types of \ac{VLSI} learning systems. Additional experimental
results from the circuits shown in Fig.~\ref{fig:learn} are presented
in Section~\ref{sec:experimental-results}.

\section{From circuits to networks} %
\label{sec:networks}

\begin{figure}
  \centering
  \includegraphics[width=0.5\textwidth]{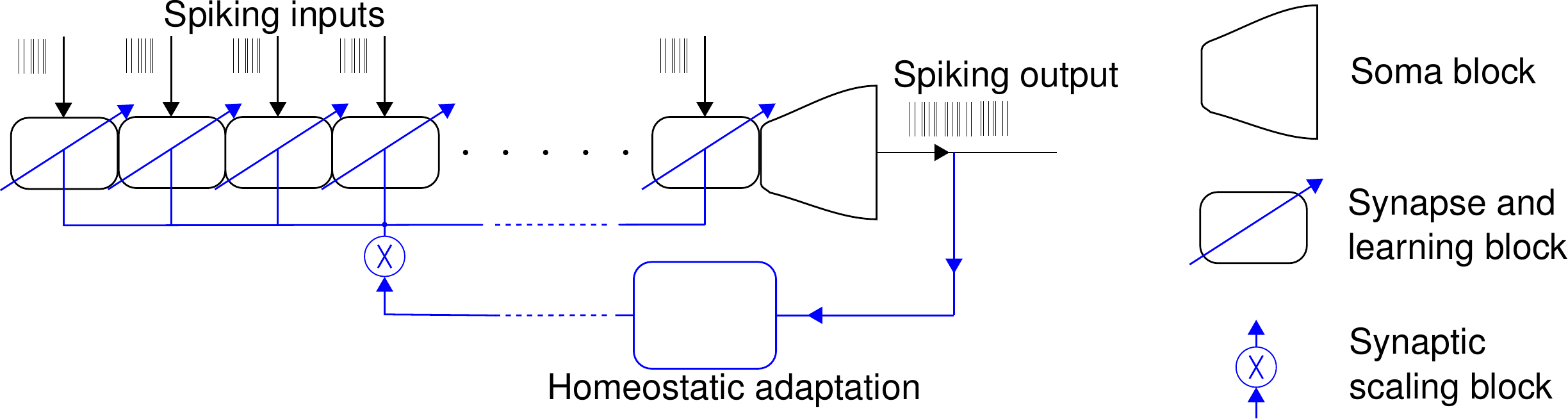}
  \caption{Silicon neuron diagram. This is a schematic representation
    of a typical circuital block comprising multiple synapse blocks,
    an \ac{IF} soma block, and a homeostatic plasticity control
    block. The synapses receive input spikes, integrate them and and
    convey the resulting currents to the soma. The soma integrates
    these currents and produces output spikes with a mean rate that is
    proportional to the total net input current.  Synapse circuits can
    implement both local plasticity mechanisms to change their
    efficacy, and global scaling mechanisms via additional homeostatic
    control block.}
  \label{fig:neurdiagram}
\end{figure}

\begin{figure}
  \centering
  \begin{subfigure}[b]{0.45\textwidth}
    \centering
    \includegraphics[width=0.7\textwidth]{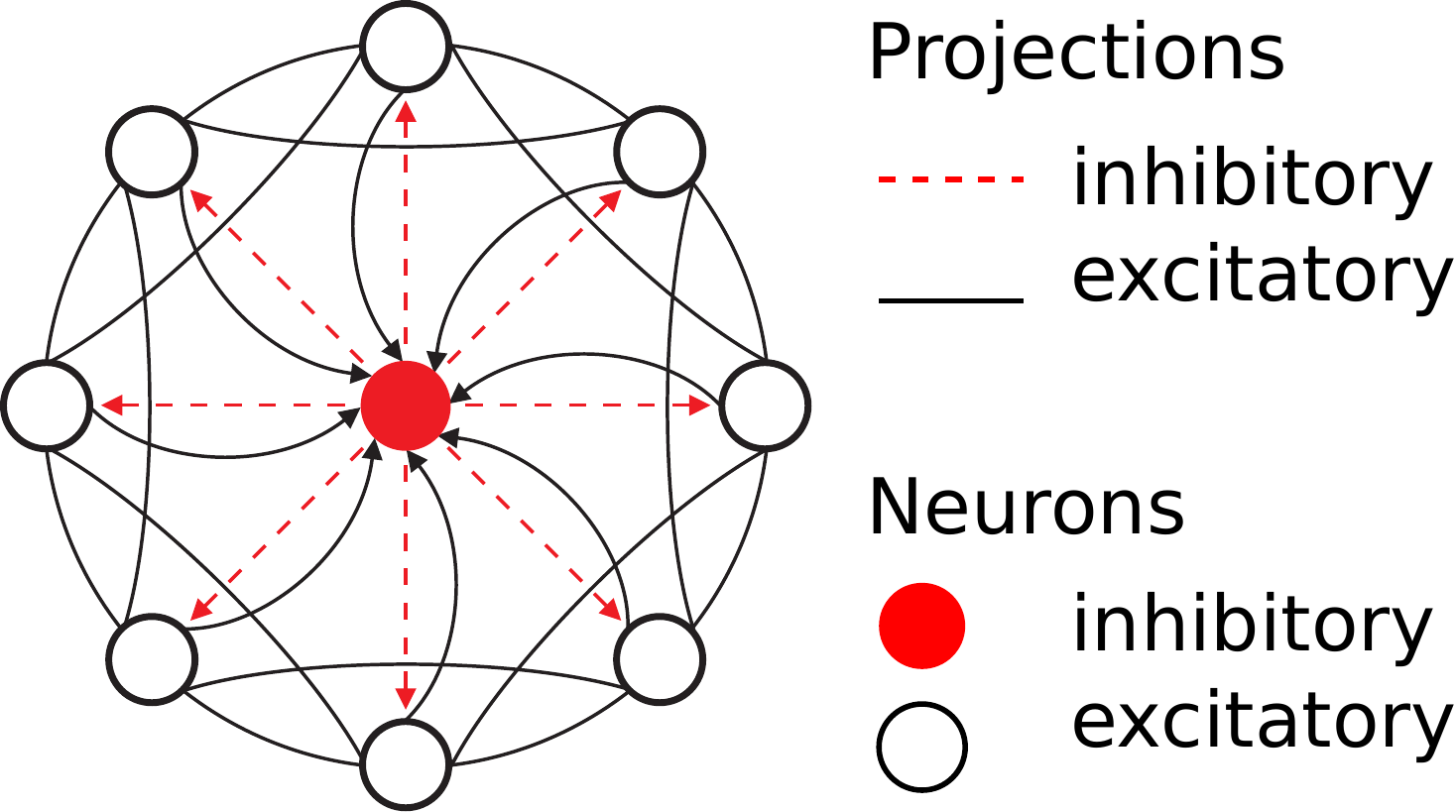}
    \subcaption{}
    \label{fig:arch1}
  \end{subfigure}\\
    \vspace{1em}
  \begin{subfigure}[b]{0.45\textwidth}
    \includegraphics[width=0.9\textwidth]{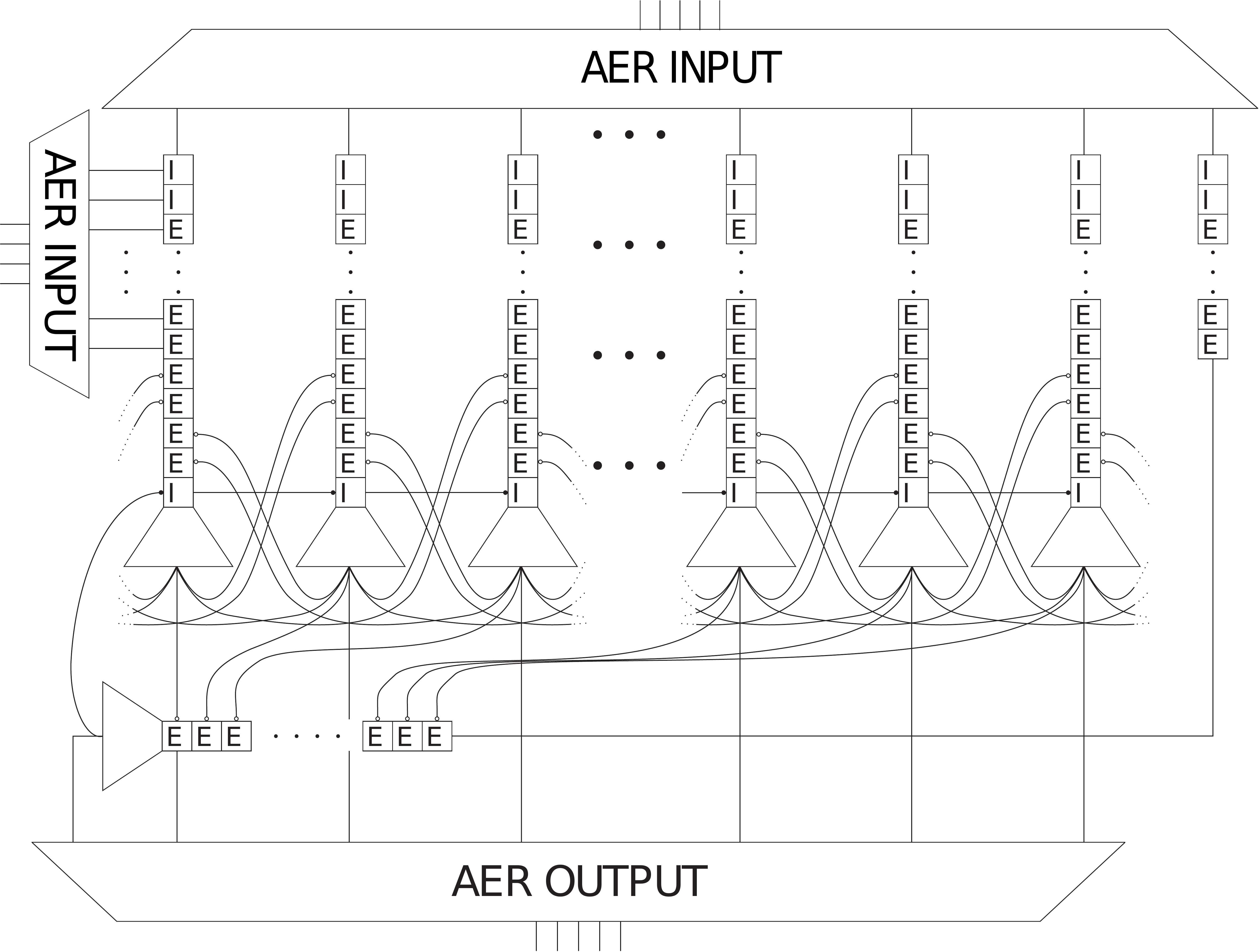}
    \subcaption{}
    \label{fig:arch2}
  \end{subfigure}
  \caption{\ac{sWTA} network topology. (\subref{fig:arch1}) Schematic
    representation of the connectivity pattern of the \ac{sWTA}
    network. These connections are implemented by synapses with
    hardwired connections to pre- and post-synaptic neurons. Empty
    circles represent excitatory neurons and the filled circle
    represents the global inhibitory neuron. Solid/dashed lines
    represent excitatory/inhibitory connections.  Connections with
    arrowheads are mono-directional, all the others are bidirectional.
    Only 8 excitatory neurons are shown for simplicity.
    (\subref{fig:arch2}) Chip architecture. Squares represent
    excitatory (E) and inhibitory (I) synapses, small unlabeled
    trapezoids represent \ac{IF} neurons. The \ac{IF} neurons transmit
    their spikes off-chip and/or to locally connected synapses implementing the
    network topology depicted in (\subref{fig:arch1}). Adapted
    from~\cite{Chicca_etal07}.}
  \label{fig:arch}
\end{figure}

The silicon neuron, synapse, and plasticity circuits presented in the
previous Sections can be combined together to form full networks of
spiking neurons. Typical spiking neural network chips have the
elements described in Fig.~\ref{fig:neurdiagram}. Multiple instances
of these elements can be integrated onto single chips and connected
among each other either with on-chip hard-wired connections (e.g., see
Fig.~\ref{fig:arch1}), or via off-chip reconfigurable connectivity
infrastructures~\cite{Fasnacht_Indiveri11, Scholze_etal11,
  Fasnacht_etal08, Chicca_etal07b, Gomez-Rodriguez_etal06}.

\subsection{Recurrent neural networks}
\label{sec:soft-wta}

In the most general \ac{RNN} each neuron is connected to every other
neuron (fully recurrent network). Unlike feed-forward networks, the
response of \acp{RNN} to the input does not only depend on the
external input but also on their internal dynamics, which in turn is
determined by the connectivity profile. Thus, specific changes in
connectivity, for example through learning, can tune the \ac{RNN}
behavior, which corresponds to the storage of internal
representations of different external stimuli. This property makes \acp{RNN}
suitable for implementing, among other properties, associative
memories~\cite{Amit_Fusi94}, working memory~\cite{Mongillo_etal03},
context-dependent decision making~\cite{Rigotti_etal10}.

There is reason to believe that, despite significant variation across cortical
areas, the pattern of connectivity between cortical neurons is similar
throughout neocortex. This fact would imply that the remarkably wide range of
capabilities of the cortex are the results of a specialization of different
areas with similar structures to the various
tasks~\cite{Douglas_etal89, Douglas_Martin04}. An intriguing hypothesis about
how computation is carried out by the brain is the existence of a finite set of
computational primitives used throughout the cerebral cortex. If we could
identify these computational primitives and understand how they are implemented
in hardware, then we would make a significant step toward understanding how to
build brain-like processors. There is an accumulating body of evidence that
suggests that one potential computational primitive consists of a \ac{RNN} with
a well defined excitatory/inhibitory connectivity
pattern~\cite{Douglas_Martin04} typically referred as \ac{sWTA} network.

In \ac{sWTA} neural networks, group of
neurons compete with each other in response to an
input stimulus. The neurons with highest response suppress all other
neurons to win the competition. Competition is achieved through a
recurrent pattern of connectivity involving both excitatory and
inhibitory connections. Cooperation between neurons with similar
response properties (e.g., close receptive fields or stimulus
preference) is mediated by excitatory connections.
Competition and cooperation make the output of an individual neuron
depend on the activity of all neurons in the network and not just on
its own input~\cite{Douglas_etal95}. As a result, \acp{sWTA} perform
not only common linear operations but also complex non-linear
operations~\cite{Douglas_Martin07}. The linear operations include
analog gain (linear amplification of the feed-forward input, mediated
by the recurrent excitation and/or common mode input), and locus
invariance~\cite{Hansel_Sompolinsky98}. The non-linear operations
include non-linear selection~\cite{Amari_Arbib77, Dayan_Abbott01,
  Hahnloser_etal00}, signal restoration~\cite{Dayan_Abbott01,
  Douglas_etal95b}, and multi-stability~\cite{Amari_Arbib77,
  Hahnloser_etal00}.

The computational abilities of these types of networks are of great
importance in tasks involving feature-extraction, signal restoration
and pattern classification problems~\cite{Maass00}. For example,
localized competitive interactions have been used to detect elementary
image features (e.g., orientation)~\cite{Ben-Yishai_etal95,
  Somers_etal95}. In these networks, each neuron represents one
feature (e.g., vertical or horizontal orientation); when a stimulus is
presented the neurons cooperate and compete to enhance the response to
the features they are tuned to and to suppress background noise.  When
\ac{sWTA} networks are used for solving classification tasks, common
features of the input space can be learned in an unsupervised
manner. Indeed, it has been shown that competition supports
unsupervised learning because it enhances the firing rate of the
neurons receiving the strongest input, which in turn triggers learning
on those neurons~\cite{Bennett90}.

\subsection{Distributed multi-chip networks}
\label{sec:distr-multi-chip}

The modularity of the cortex described in the theoretical works and
suggested by the experimental observations above mentioned,
constitutes a property of great importance related to the scalability
of the system. If we understood the principles by which such
computational modules are arranged together and what type of
connectivity allows for coherent communication also at large
distances, we would be able to build scalable systems, i.e., systems
whose properties are qualitatively reproduced at all scales.

The idea of modularity poses some technological questions as to how
the communication between the systems should be implemented.  Large
\ac{VLSI} networks of \ac{IF} neurons can already be implemented on
single chips, using today's technology.  However implementations of
pulse-based neural networks on multi-chip systems offer greater
computational power and higher flexibility than single-chip systems
and constitute a tool for the exploration of the properties of
scalability of the neuromorphic systems.  Because inter-chip
connectivity is limited by the small number of input-output
connections available with standard chip packaging technologies, it is
necessary to adopt time-multiplexing schemes for constructing large
multi-chip networks. This scheme should also allow for an asynchronous
type of communication, where information is transmitted only when
available and computation is performed only when needed in a
distributed, non-clocked manner.

In recent years, we have witnessed the emergence of a new asynchronous
communication standard that allows analog \ac{VLSI} neurons to
transmit their activity across chips using pulse-frequency modulated
signals (in the form of events, or spikes). This standard is based on
the \ac{AER} communication protocol~\cite{Mahowald92}. In \ac{AER}
input and output signals are real-time asynchronous digital pulses
(events or spikes) that carry analog information in their temporal
relationships (inter-spike intervals).  If the activity of the
\ac{VLSI} neurons is sparse and their firing rates are biologically
plausible (e.g., ranging from a few spikes per second to a few hundred
spikes per second), then it is possible to trade-off space with speed
very effectively, by time-multiplexing a single (very fast) digital
bus to represent many (very slow) neuron axons. For example, it has
been recently demonstrated how these time-multiplexing schemes can
sustain more then $60$\,M~events/sec, representing the
\emph{synchronous} activity of $1$\,M neurons firing at a rate of
$60$\,Hz~\cite{Fasnacht_Indiveri11, Boahen00}. In general, \ac{AER}
communication infrastructures provide the possibility to implement
arbitrary custom multi-chip architectures, with flexible connectivity
schemes. Address events can encode the address of the sending node
(the spiking neuron) or of the receiving one (the destination
synapse). The connectivity between different nodes is implemented by
using external digital components and is typically defined as a
look-up table with source and destination pairs
  of addresses, or by more resource-efficient schemes e.g., using
  multicast or multi-stage routing~\cite{Painkras_etal13,
    Carrillo_etal12,Moradi_etal13}.
This asynchronous digital solution permits flexibility in the configuration
(and re-configuration) of the network topology, while keeping the computation
analog and low-power at the neuron and synapse level.

To handle cases in which multiple sending nodes attempt to transmit
their addresses at exactly the same time (event collisions), on-chip
digital asynchronous arbitration schemes have been
developed~\cite{Mahowald92, Boahen00, Boahen04}. These circuits work
by queuing colliding events, so that only one event is transmitted at
a time. Multiple colliding events are therefore delayed by a few
nano-seconds or fractions of microseconds. For neuromorphic
architectures that use biologically plausible time constants (i.e., of
the order of milliseconds), these delays are negligible and do not
affect the overall performance of the network. For example, assuming a
tolerance of $1$\,ms jitter~\cite{Hatsopoulos_etal03} it is possible to
process up to $4$\,K coincident input events without introducing
sensible delays, even with an outdated $350$\,nm \ac{CMOS}
technology~\cite{Chicca_etal07b}. On the other hand, in
accelerated-time systems, such as those proposed
in~\cite{Pfeil_etal13} whose circuits operate at $10^4$ the speed of
their biological counterpart, communication delays are much more
critical, because their duration does not scale. In general, the
performance of any \ac{AER} neuromorphic system will be bound by
communication memory and bandwidth constraints, which trade-off the
speed of the neural processing elements with the size of the network
that can be implemented.

\subsection{A \ac{SW}/\ac{HW} echo-system}
\label{sec:acswachw-echo-system}

In order to promptly explore the computational properties of different
types of large-scale multi-chip computational architectures,
it is important to develop a dedicated \ac{HW} and \ac{SW}
infrastructure, which allows a convenient, user-friendly way to
define, configure, and control in real-time the properties of the
\ac{HW}~\cite{Davison_etal08,Sheik_etal11} spiking neural networks,
as well as a way to monitor in real-time their spiking and non-spiking
activity.

The definition of a \ac{SW} infrastructure for
neuromorphic systems pertains to an issue of increasing importance. Indeed, as
reconfigurable neuromorphic platforms are scaled to larger sizes, it is
necessary to develop efficient tools to interpret the neural network model,
e.g., through programming or scripting languages, and configure the hardware
parameters correspondingly for the neural and synaptic dynamics and for the
events routing. Hence, the \ac{SW} should provide means to configure, control,
interact and monitor the electronic hardware. Fortunately, while the specific
electronic implementation of each neuromorphic system can differ substantially, several
common properties can be identified, such as the use of an \ac{AER} scheme for
communication. Therefore a \ac{SW} \emph{echo-system} can be defined to assemble
and control the system in a modular, fully reconfigurable way. In this respect,
several \ac{SW} interfaces for neuromorphic and neuro-computing platforms have
already been developed. The scopes of these tools are diverse and so are
their peculiarities due to the specificities of the corresponding system. Both
digital neuro-computing platforms and analog neuromorphic systems typically
require a ``neuromorphic compiler'' able to parse the network topology and
configure correspondingly memories, processors or digital interfaces to
properly simulate the neural and synaptic dynamics and route the spiking events
through the network~\cite{Patterson_etal12, Galluppi_etal12b, Minkovich_etal12,
Preissl_etal12}. On top of the compilers, a number of \ac{SW} frameworks have
been developed as scripting and programming languages for neural networks at
the level of the single network elements, e.g., neurons, synapses and
connectivity~\cite{Davison_etal08} and also including a system-level
description for building large-scale, brain simulators~\cite{Stewart_etal09}.

A promising example of an open-source \ac{SW} framework that
interprets generalized hardware specification files and constructs an
abstract representation of the neuromorphic devices compatible with
high-level neural network programming libraries is available
at~\url{http://inincs.github.com/pyNCS/}. This framework is based on
reconfigurable and extensible \acp{API} and includes a high-level
scripting front-end for defining neural networks. It constitutes a
bridge between applications using abstract resources (i.e., neurons
and synapses) and the actual processing done at the hardware level
through the management of the system's resources, much like a
\emph{kernel} in modern computers~\cite{Wulf_etal74}, and it is
compatible with most existing software.  The \ac{HW} and \ac{SW}
infrastructure can be complemented with tools for dynamic parameter
estimation methods~\cite{Neftci_etal11, Neftci_etal12} as well as
automated methods for measuring and setting circuit-level parameters
using arbitrary cost-functions at the network
level~\cite{Sheik_etal11}.

\section{Experimental results}   %
\label{sec:experimental-results} %

The circuits and architectures described in this paper have been
designed and developed over the course of several years.  Therefore
the experimental data presented in this Section has been collected
from multiple neuromorphic \ac{VLSI} devices and systems. The results
presented demonstrate the correct behavior of the circuits described
in the previous Sections.

\subsection{Synaptic and neural dynamics}
\label{sec:synapt-neur-dynam}

\begin{figure}
  \centering
  \includegraphics[width=0.4\textwidth]{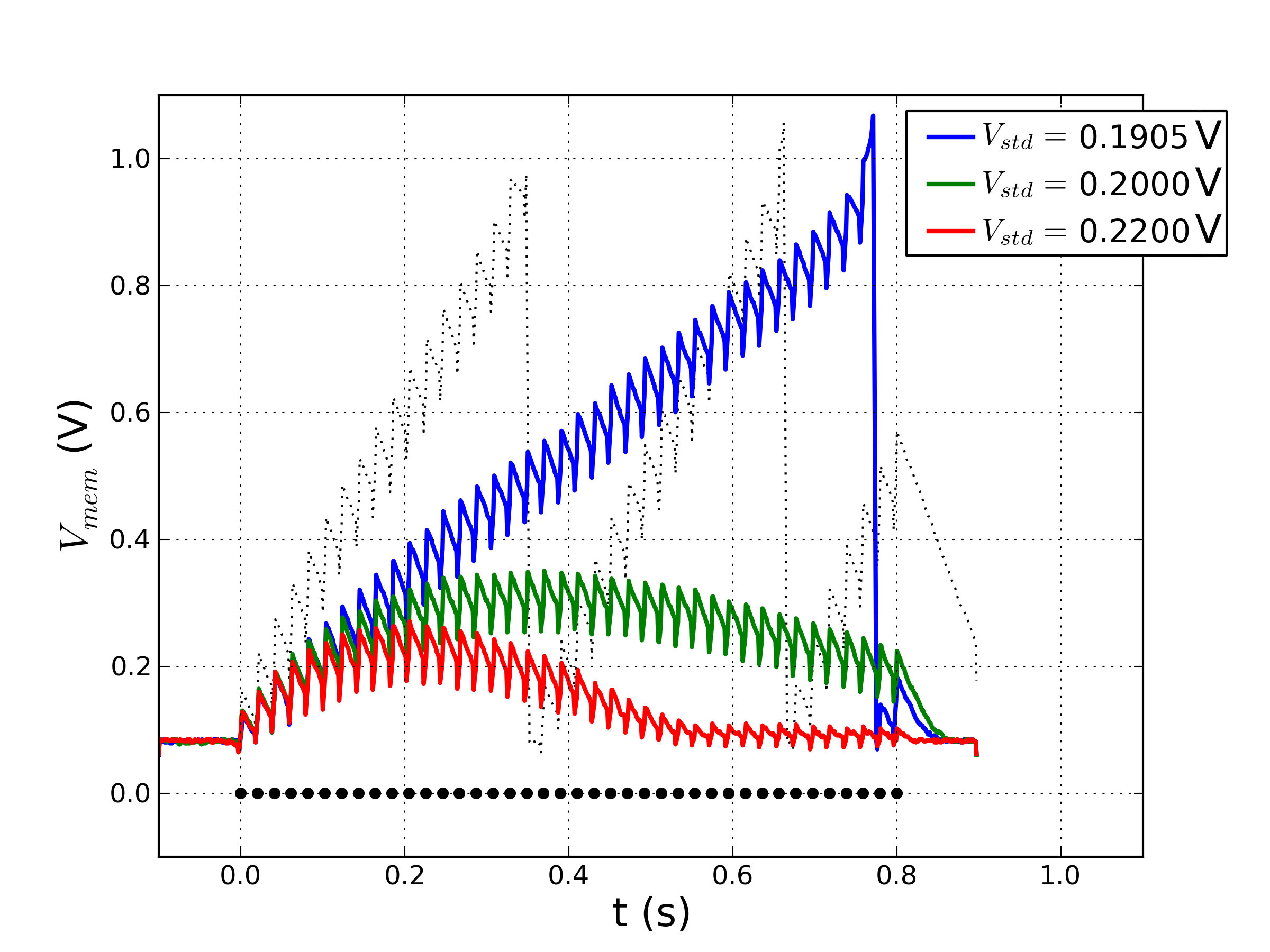}
  \caption{Membrane potential of \ac{IF} neuron in response to a
    $50$\,Hz pre-synaptic input spike train for different values of
    short-term depression adaptation rate, which is controlled by $V_{std}$
    bias (see Fig.~\ref{fig:dpi-tot}). The dashed trace in
    background corresponds to the response without \ac{STD}. Black
    dots correspond to input spike-times.}
  \label{fig:std}
\end{figure}

To show the combined effect of synaptic and neural dynamics, we
stimulated a silicon neuron via an excitatory \ac{DPI} synapse
circuit, while sweeping different \ac{STD} parameter settings.  The
typical phenomenology of \ac{STD} manifests as a reduction of
\ac{EPSC} amplitude with each presentation of a pre-synaptic spike,
with a slow (e.g., of the order of $100$\,ms) recovery
time~\cite{Markram_Tsodyks96}. 
In Fig.~\ref{fig:std} we plot the neuron's membrane potential
$V_{mem}$ during the stimulation of one of its excitatory synapses
with a regular pre-synaptic input spike train of $50$\,Hz, for
different \ac{STD} adaptation settings. Small parameter settings for
the \ac{STD} bias voltage have no or little effect. But for larger
settings of this bias voltage the effect of \ac{STD} is prominent: the
synaptic efficacy decreases with multiple input spikes to a point in
which the net input current to the soma becomes lower than the
neuron's leak current, thus making the neuron membrane potential
decrease, rather than increase over time.

\begin{figure}
  \centering
  \includegraphics[width=0.375\textwidth]{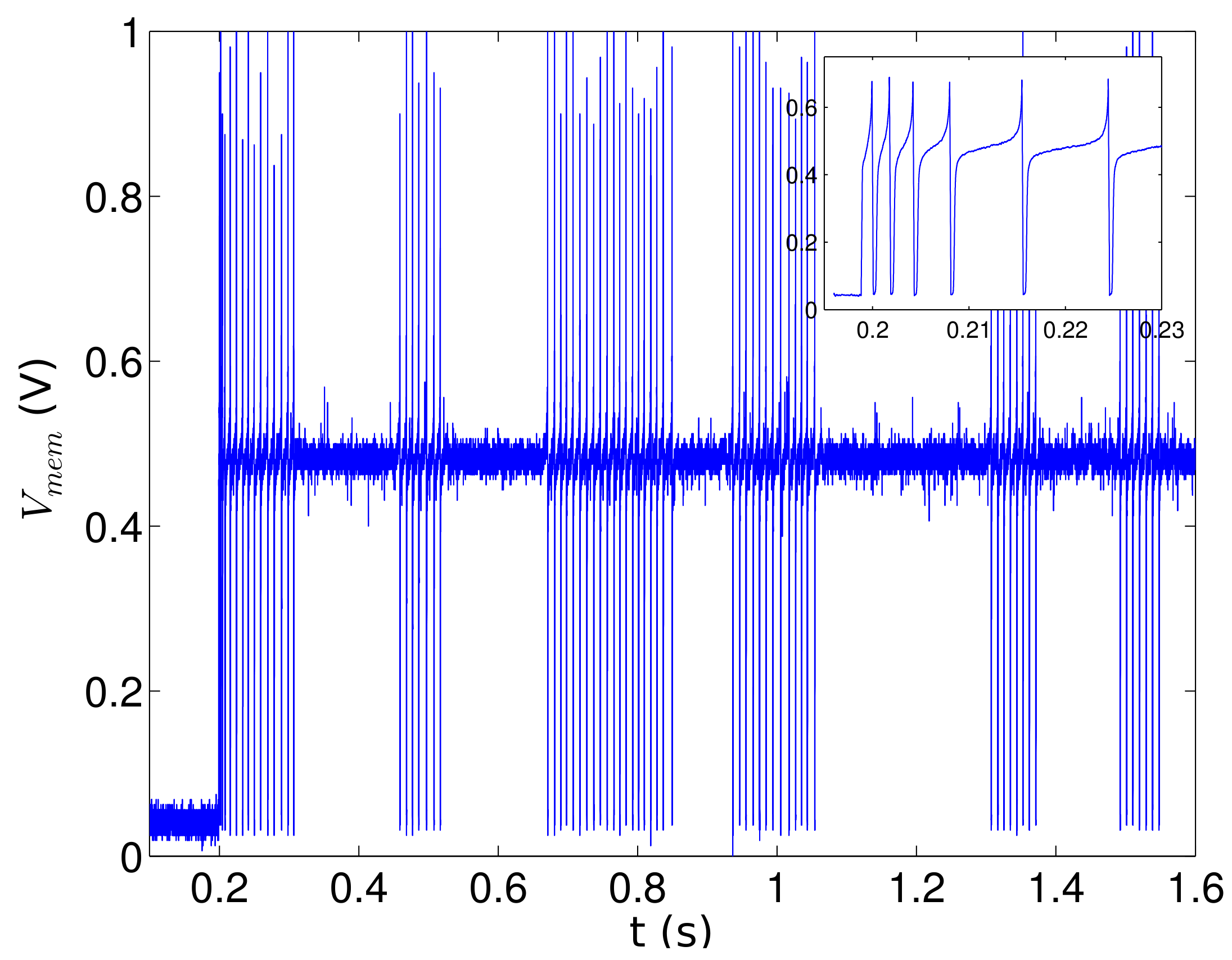}
  \caption{Silicon neuron response to a step input current, with spike
    frequency adaptation mechanism enabled and parameters tuned to
    produce bursting behavior. The figure inset represents a zoom of
    the data showing the first 6 spikes. Adapted from~\cite{Indiveri_etal10}.}
  \label{fig:adap}
\end{figure}

Another important adaptation mechanism discussed in
Section~\ref{sec:silicon-neurons}, is that of \emph{spike-frequency
  adaptation}. To show the effect of this mechanism, we set the
relevant bias voltages appropriately, stimulated the silicon neuron
with a constant input current, and measured it's membrane potential.
Figure~\ref{fig:adap} shows an example response to the step input
current, in which $V_{lkahp}=0.05\,V$, $V_{thrahp}=0.14\,V$,
$V_{ahp}=2.85\,V$. As shown, we were able to tune the adaptation
circuits in a way to produce bursting behavior. This was achieved by
simply increasing the gain of the negative feedback adaptation
mechanism ($V_{thrahp}>0$). This is equivalent to going from an
asymptotically stable regime to a marginally stable one, that produces
ringing in the adaptation current $I_{ahp}$, which in turn produces
bursts in the neuron's output firing rate. This was possible due to
the flexibility of the \ac{DPI} circuits, which allow us to take
advantage of the extra control parameter $V_{thrahp}$, in addition to
the adaptation rate parameter $V_{ahp}$, and the possibility of
exploiting its non-linear transfer properties as described in
Section~\ref{sec:silicon-synapses}, without requiring extra circuits
or dedicated resources that alternative neuron models have to
use~\cite{Folowosele_etal09,Mihalas_Niebur09,Wijekoon_Dudek08}.

\subsection{Spike-based  learning} %
\label{sec:learning-circuits}

In this section we present measurements from the circuits implementing
the \ac{STDP} learning mechanism described in
Section~\ref{sec:lcircuits}. To stimulate the synapses we generated
pre-synaptic input spike trains with Poisson distributions. Similarly,
the post-synaptic neuron was driven by a current produced via a
non-plastic synapse (a \ac{DPI} circuit with a constant synaptic
weight bias voltage) stimulated by software-generated Poisson spike
trains.  These latter inputs are used to drive the \ac{IF} neuron
towards different activity regimes which regulate the probabilities of
synaptic transitions~\cite{Fusi_Mattia99,Brader_etal07}, effectively
modulating the learning rate in unsupervised learning
conditions, or acting as teacher signals in supervised learning
conditions.

\begin{figure}
  \centering
  \includegraphics[width=0.4\textwidth]{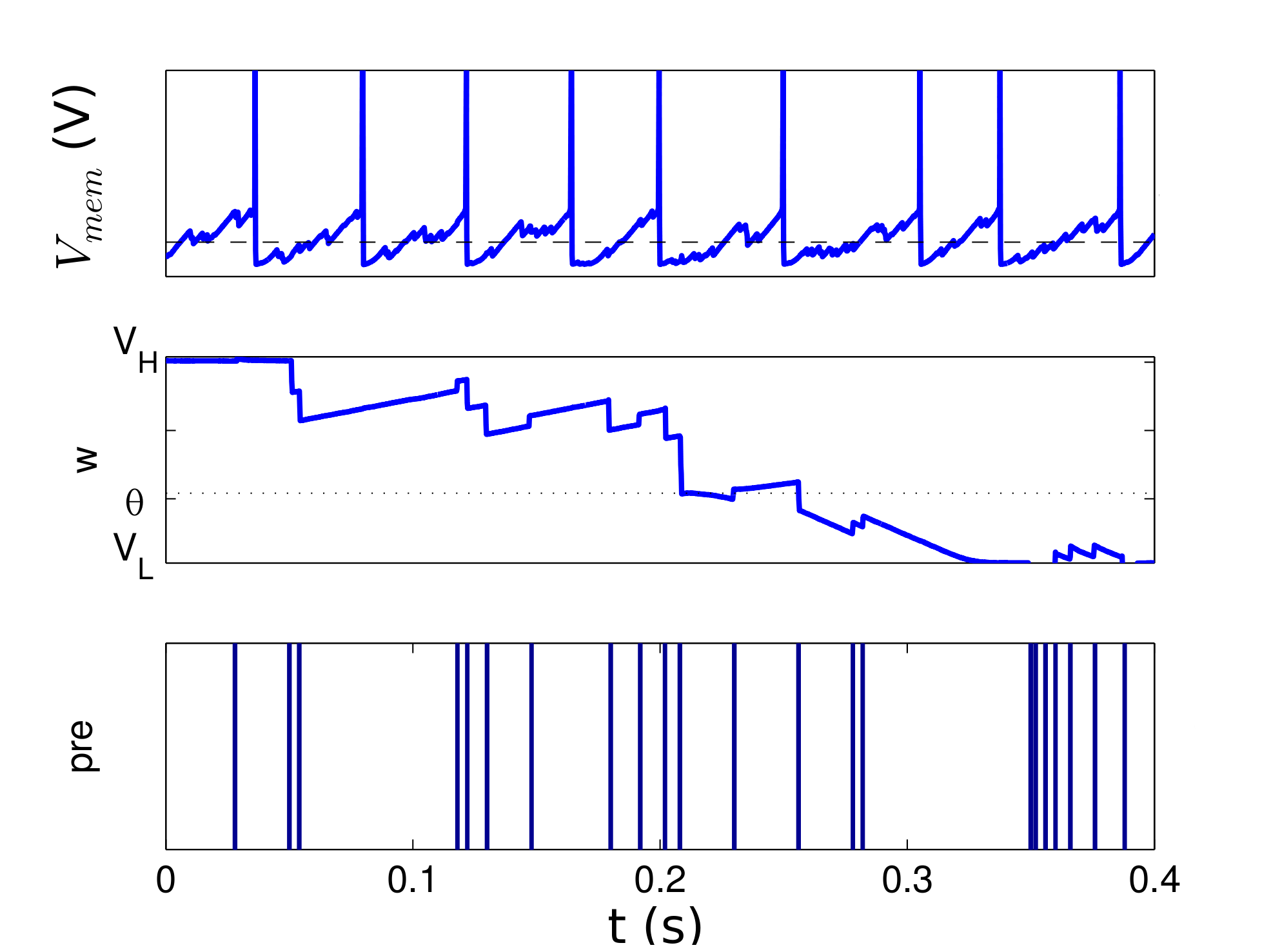}
  \caption{Stochastic transitions in synaptic states. The non-plastic synapse
  is stimulated with a Poisson distributed spikes train. The neuron fires at an
  average rate of $~30$\,Hz. The pre-synaptic input ($V_{pre}$) is stimulated
  with Poisson distributed spike trains with a mean firing rate of $60$\,Hz. The
  updates in the synaptic weight produced an \acs{LTD} transition that remains
  consolidated. $V_H$ and $V_L$ show the potentiated and depressed levels
  respectively while $w$ denotes the synaptic weight, and $\theta$ the
  bi-stability threshold. Adapted from~\cite{Mitra_etal09}.}
  \label{fig:lUD}
\end{figure}

The Poisson nature of the spike-trains used in this way represents the
main source of variability required for implementing stochastic
learning~\cite{Fusi_etal00, Chicca_Fusi01}. In Fig.~\ref{fig:lUD} we show
measurements from a stochastic learning experiment in which the neuron is
driven to a regime where both potentiation and depression are possible but
depression has a higher probability to occur. As shown, the weight voltage
undergoes both positive and negative changes, depending on the timing of the
input spike and the state of the post-synaptic neuron (as explained in
Section~\ref{sec:lcircuits}). In addition, the weight voltage is slowly driven
toward one of the two stable states, depending on whether it is above or below
the threshold $\theta$ (where $\theta$ corresponds to the voltage $V_{thw}$ of
Fig.~\ref{fig:lsyn}). Long-term transitions occur when a series of pre-synaptic
spikes arrive in a short time-frame causing the weight to cross the threshold
$\theta$. As a consequence, the probability of synaptic state transitions
depends on the probability that such events occur, hence it depends on the
firing rate of the pre-synaptic neuron~\cite{Fusi02, Indiveri_etal06}. In the
case of the experiment of Fig.~\ref{fig:lUD} an \ac{LTD} transition has
occurred upon the presentation of an input stimulus of $60$\,Hz for $400$\,ms.
In conclusion, the bi-stability of the synapses and the spike-based plasticity
concur in a mechanism that (1) ensures that only a random fraction of the
stimulated bi-stable synapses undergo long-term modifications and (2) that
synaptic states are resilient to changes due to spontaneous activity, thus
increasing the robustness to noise.

\begin{figure}
  \centering
  \begin{subfigure}[b]{0.4\textwidth}
    \includegraphics[width=\textwidth]{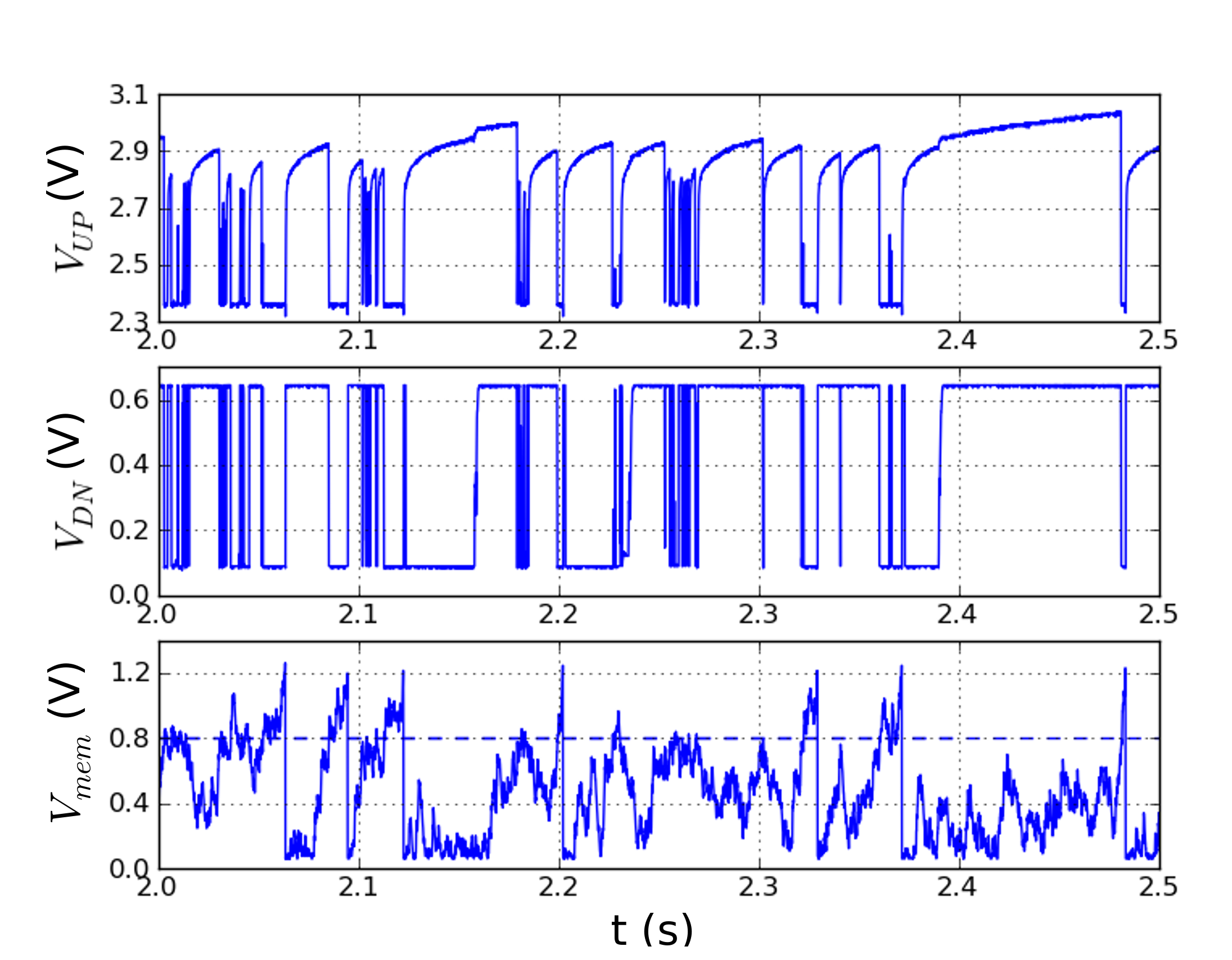}
    \subcaption{}
    \label{fig:lupdn_low}
  \end{subfigure}\\
    \vspace{1em}
  \begin{subfigure}[b]{0.18\textwidth}
    \centering
    \includegraphics{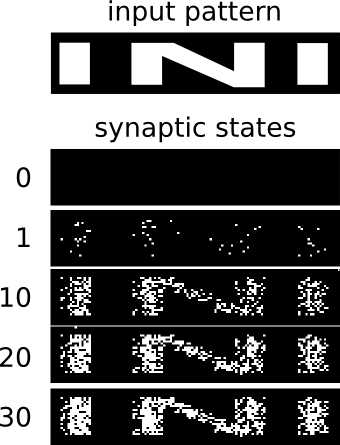}
    \subcaption{}
    \label{fig:lwijini}
  \end{subfigure}
  \begin{subfigure}[b]{0.275\textwidth}
    \centering
    \includegraphics[width=\textwidth]{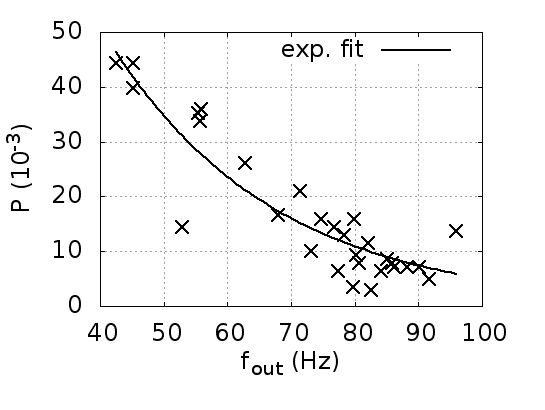}
    \subcaption{}
    \label{fig:deltasvsrates}
  \end{subfigure}
  \caption{Stochastic learning.  (\subref{fig:lupdn_low}) Single
    neuron stochasticity. Traces from a VLSI multi-neuron chip with
    I\&F neurons and plasticity circuits as in
    Fig.~\ref{fig:lsyn}. The $V_{UP}$ and $V_{DN}$ signals (top
    traces) are set by the circuits in Fig.~\ref{fig:lsoma}.  A
    Poisson spike-train of high firing rate is sent to the excitatory
    synapse of a I\&F neuron whose $V_{mem}$ trace is reported in the
    lower trace. The strong input current generated by the synapse has
    been compensated by a strong leakage current
    ($V_{leak}=0.39$\,V). This parameter choice allows to exploit the
    stochasticity of the input spike-trains to produce the highly
    irregular dynamics of $V_{mem}$. The non-ideal rounding in the
    rising part of the $V_{UP}$ trace has negligible effects on the
    synaptic weight given the exponential nature of the current
    generated through transistor $M_{L3}$ of Fig.~\ref{fig:lsyn}.
    (\subref{fig:lwijini}) An image of the ``INI'' acronym is
    converted into a series of Poisson spike-trains and gradually
    stored in the memory by repeated presentations. See text for
    details.  (\subref{fig:deltasvsrates}) Normalized frequency of
    occurrence of \acs{LTP} transitions during the experiment of
    Fig.~\subref{fig:lwijini}, fitted by an exponential function
    (dashed line).}
    \label{fig:learnupdn}
    \end{figure}

If Fig.~\ref{fig:lupdn_low} we show the results of another stochastic learning
experiment in which we stimulated the post-synaptic neuron with a
high-frequency Poisson-like spike train through a non-plastic excitatory input
synapse, in order to produce Poisson-like firing statistics in the output. The
dashed line on the $V_{mem}$ plot represents the learning threshold voltage
$V_{thm}$ of Fig.~\ref{fig:lsoma}. The $V_{UP}$ (active low) and $V_{DN}$
(active high) signals are the same shown in Fig.~\ref{fig:lsoma} and represent
the currents that change the synaptic values when triggered by pre-synaptic
spikes. They can be considered as eligibility traces that enable the weight
update mechanism when they are active.

In Fig.~\ref{fig:lwijini} we show the results of an experiment where
we trained a matrix of $28\times124 = 3472$ plastic synapses,
constituting the total input of a neuron, with multiple presentations
of the same input pattern representing the ``INI'' acronym. Initially
all the neuron's input synaptic weights are set to their low state
(black pixels). Then, the post-synaptic neuron is driven by a teacher
signal that makes it fire stochastically with a mean rate of
$40$\,Hz. At the same time, input synapses are stimulated according to
the image pattern: in the input image (top left image), each white
pixel represents a Poisson spike train of $55$\,Hz, sent to the
corresponding synapse; similarly, each black pixel represents a low
rate spike train ($5$\,Hz) which is transmitted to its corresponding
synapse. Because the probability of \ac{LTP} depends on the
pre-synaptic firing rate, elements of the input matrix that correspond
to a white pixel have are more likely to make a transition to the
potentiated state compared to the other ones. Because of the
stochastic nature of the input patterns, only a random subset of
synapses undergoes \ac{LTP}, leaving room available to store other
memories. By repeating the presentation of the input pattern multiple
times, this pattern gets gradually \emph{stored} in the synaptic
matrix.  The bottom left image of Fig.~\ref{fig:lwijini} represents
the synaptic matrix at the end of the experiment. Furthermore, the
\emph{stop-learning} mechanism described in Sec.~\ref{sec:lcircuits}
causes a drop in the number of synapses that undergo \ac{LTP} because
as the pattern is stored in the memory the post-synaptic firing rate
increases (Fig.~\ref{fig:deltasvsrates}).

The above experiments demonstrate the properties of the learning
circuits implemented in the VLSI chips. In a feed-forward
configuration, the neuron can be controlled by an external spiking
teacher signal, which indirectly controls the transition
probabilities. This ``perceptron-like'' configuration allows the
realization of supervised learning protocols for building real-time
classification engines. But, as opposed to conventional
perceptron-like learning rules, the spike-triggered weight updates
implemented by these circuits overcome the need for an explicit
control (e.g., using error back-propagation) on every individual
synapse. In ``Hopfield-network'' like \ac{RNN} configurations the same
neuron and plasticity circuits can implement \ac{ANN}
learning schemes~\cite{Giulioni_etal08,Giulioni_etal12}, exploiting
the neural network dynamics to form memories through stochastic
synaptic updates, without the need for explicit random generators at
each synapse.

\subsection{\ac{sWTA} networks of \ac{IF} neurons}
\label{sec:netw-acif-neur}

\begin{figure}
  \centering
  \includegraphics[width=0.375\textwidth]{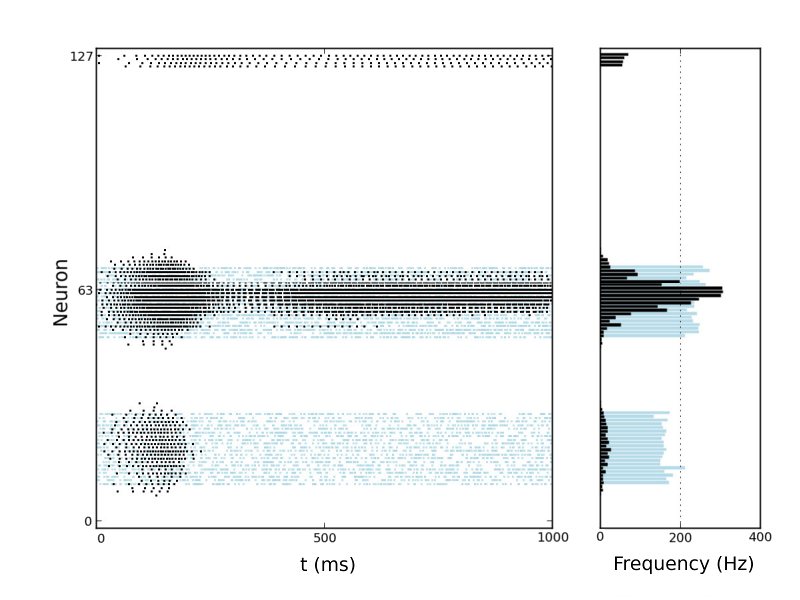}
  \includegraphics[width=0.375\textwidth]{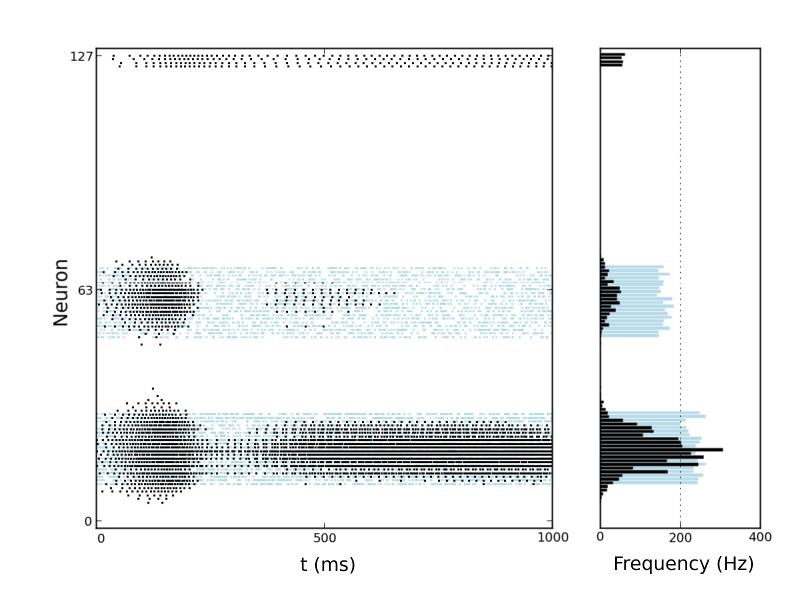}
  \caption{Selective amplification experiments. The network is
    stimulated in two regions, one centered around unit 20 and the
    other around unit 60, with Poisson spike trains of mean firing
    rate $180$\,Hz and $240$\,Hz.  The figures show the networks
    response to these inputs (black) and their respective steady state
    firing rates on the right panels (calculated for time $>\
    500$\,ms). Neurons 124 to 127 are the 4 inhibitory neurons of the
    soft \acs{WTA} network.  In the right and left panel the input
    amplitudes are swapped. The results show smooth activity profiles
    that are invariant to input swapping, demonstrating that the
    mismatch in the local weights has been partially compensated.
    Adapted from~\cite{Neftci_Indiveri10}.}
  \label{fig:wta-data}
\end{figure}

Two characteristic features of \ac{sWTA} networks that make them ideal
building blocks for cognitive systems are their ability to selectively
enhance the contrast between localized inputs and to exhibit activity
that persists even after the input stimulus has disappeared. We
configured the local hardwired connectivity of a multi-neuron chip to
implement an \ac{sWTA} network and carried out test experiments to
show both selective amplification and state-dependent
computation. Specifically, we configured a chip comprising a network
of 128 \ac{IF} neurons with local nearest neighbor excitatory
connectivity and global inhibition: each neuron was configured to
excite its first nearest neighbors, its second neighbors and a
population of four global inhibitory neurons (the top four neurons in
the array of 128 neurons). In a first experiment, we calibrated the
settings and input stimuli to minimize the effect of device mismatch,
following the event-based techniques described
in~\cite{Sheik_etal11,Neftci_etal11} and stimulated the network with
two distinct regions of activation, centered around units 20 and 60
(see shaded areas in Fig.~\ref{fig:wta-data}). In one case the top
region had a higher mean firing rate than the bottom one and in the
other case the bottom region had a higher activation (see top and
bottom plots in Fig.~\ref{fig:wta-data} respectively). As expected
from
theory~\cite{Hansel_Sompolinsky98,Dayan_Abbott01,Douglas_Martin07},
the population of silicon neurons receiving the strongest input won
the competition, enhancing its activity by means of the local
recurrent connections, while suppressing the activity of the competing
population via the global inhibitory connections (selective
amplification feature).

In a second experiment we demonstrate the behavior of a \ac{sWTA}
architecture used to construct state-holding elements, which are the
basic blocks for building \acp{FSM} using spiking neurons, and in
which the \ac{FSM} states are represented by sub-populations of
neurons. The network topology supporting the \ac{FSM} functionality
and used in the following experiments resembles the ones of \ac{ANN}
with discrete or line-attractors. As mentioned in the previous
sections, this type of networks can support a diverse range of
functionalities and have employed in hardware implementation, e.g.,
for head-direction tracking~\cite{Massoud_Horiuchi11} and memory
recall~\cite{Giulioni_etal12}. In particular we concentrated our
experiments on demonstration of two of their main properties useful
for implementing the \ac{FSM}, namely selective amplification and
state-switching due to external inputs.

\begin{figure}
  \centering
  \includegraphics[width=0.45\textwidth]{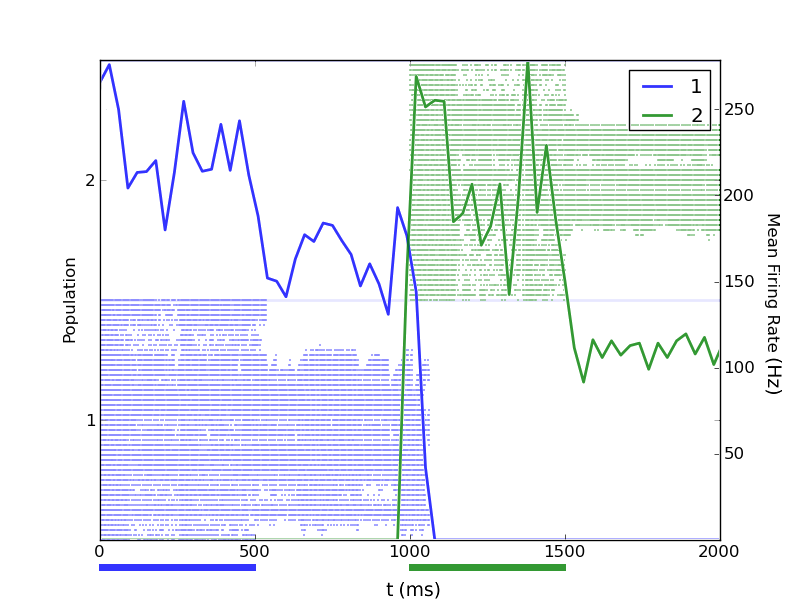}
  \caption{\acl{FSM} state-holding behavior using a \ac{VLSI}
    \ac{sWTA} architecture. States are represented by two recurrently
    connected populations of \ac{IF} neurons using the hard-wired,
    on-chip connectivity. Population 1 (bottom half of the raster
    plot) is stimulated by synthesized Poisson spike trains for the
    initial $500$\,ms. It's activity persists due to the recurrent
    excitatory connectivity, until population 2 (top half of the
    raster plot) is stimulated. The width and position of the
    sub-populations depend on the properties of the local connectivity
    and on their variability. Line-plots superimposed to the
    raster-plot represent the mean firing-rates computed across each
    population. The colored bars below the plot represent input
    stimulus presentations. Input stimuli are composed of Poisson
    spike trains of $200$\,Hz lasting for $500$\,ms, and are applied
    to all the neurons of one population.  The higher variability in
    the output, e.g., compared with Fig.~\ref{fig:wta-data}, is due to
    the absence of mismatch compensation techniques, deliberately
    omitted to highlight the differences.}
\label{fig:fsm}
\end{figure}

In this experiment we present localized and transient inputs to two
groups of neurons using synthetically generated Poisson trains (see
Fig.~\ref{fig:fsm}).  After the presentation of each input stimulus
the activity of the stimulated population persists, reverberating in
time, by means of the local recurrent excitatory connectivity. Note
that, because of the global competition, only a sub-set of the
stimulated neurons remains active.  To obtain the results shown in
Fig.~\ref{fig:fsm}, we first stimulated the bottom population for
$500$\,ms, and then after subsequent $500$\,ms we stimulated the top
population. When the second stimulus is applied a ``state transition''
is triggered: as the top population becomes active the bottom one is
suppressed. When the second stimulus is removed, the bottom population
is completely silent, and the top population remains active, in a
self-sustained activity regime.
In full \ac{FSM} systems the state transition signals would be
produced by other neuronal populations (transition populations)
responding to both incoming input stimuli and to neurons representing
the current state. A complete description and analysis of these neural
network based \acp{FSM} is presented in~\cite{Rutishauser_Douglas09},
and working examples of multi-neuron chips implementing spiking
\acp{FSM} are described in~\cite{Neftci_etal11, Neftci_etal12}.

\section{Discussion}
\label{sec:discussion}

The set of low-power hybrid analog/digital circuits presented in the
previous sections can be used as basic building blocks for
constructing adaptive fully-parallel, real-time neuromorphic
architectures. While several other projects have already developed
dedicated hardware implementations of spiking neural networks, using
analog~\cite{Wijekoon_Dudek12}, digital~\cite{Furber_Temple07,
  Cassidy_etal13} and mixed mode analog/digital~\cite{Schemmel_etal08,
  Cruz-Albrecht_etal13} approaches few~\cite{Horiuchi_Koch99,
  Boahen05, Sarpeshkar06, Hynna_Boahen09, Brink_etal13} follow the
neuromorphic approach originally proposed in the early
nineties~\cite{Mead90}.  The foundations of this neuromorphic approach
were established by pointing out that the implementation of compact
and low-power hardware models of biological systems requires the use
of transistors in the sub-threshold analog domain and the exploitation
of the physics of the \ac{VLSI} medium. We argue that the circuits and
architectures presented here adhere to this approach and can therefore
be used to build efficient biophysically realistic real-time neural
processing architectures and autonomous behaving systems.


\subsection{Device mismatch and noise}
\label{sec:device-mism-noise}
One common criticism to this sub-threshold analog VLSI design approach
is that circuits operating in this domain have a high degree of
noise. However sub-threshold current-mode circuits have lower noise
energy (noise power times bandwidth), and superior energy efficiency
(bandwidth over power) than above-threshold
ones~\cite{Sarpeshkar_etal93b,Shi09}. \revised{mismatch1}{Another
  common criticism is that device mismatch in sub-threshold circuits
  is more prominent than in above threshold circuits.  While this
  observation is correct, device mismatch is a critical problem in any
  analog \ac{VLSI} implementation of neural networks (e.g., see the
  post-calibration neuronal variability measurements of
  above-threshold accelerated time silicon neuron circuits, presented
  in~\cite{Schmuker_etal14}). In principle it is possible to minimize
  the effect of device mismatch following standard electrical
  engineering approaches and adopting appropriate analog \ac{VLSI}
  design techniques, however we argue that it is not necessary to
  adopt aggressive mismatch reduction techniques in the type of
  neuromorphic systems we propose: these }{Another common criticism is
  that sub-threshold circuits are highly affected by device
  mismatch. Indeed device mismatch effects can cause significant
  differences in the outputs of transistors that should otherwise
  produce identical currents, when they are biased in the subthreshold
  regime.  While in principle it would be possible to minimize the
  effect of device mismatch also in this domain following standard
  electrical engineering approaches and appropriate analog \ac{VLSI}
  design techniques, this is not the best strategy to follow for
  neuromorphic systems.}  techniques would lead to very large
transistor or circuit designs, which could in turn significantly
reduce the number of neurons and synapses integrated onto a single
chip (see for example~\cite{Rachmuth_etal11}, where a whole \ac{VLSI}
device was used to implement a single synapse). Rather than attempting
to \revised{reduce}{minimize}{reduce} mismatch effects using
brute-force engineering techniques at the circuit design level, the
neuromorphic engineering approach we promote in this work aims to
address these effects at the network and system level, with collective
computation, adaptation, and feedback mechanisms.
\revised{mismatch2}{For example, the plasticity mechanisms presented
  in Section~\ref{sec:lcircuits} are intrinsically robust to mismatch
  by design, and do not require precisely matched
  transistors. Moreover, it has been shown how both short- and
  long-term plasticity mechanisms can be effectively used to reduce
  the effects of device mismatch in \ac{VLSI}
  circuits~\cite{Bill_etal10,Cameron_Murray08}, and how homeostatic
  plasticity mechanisms can be used to compensate for large changes in
  the signals affecting the operation of the neurons in multi-neuron
  \ac{VLSI} systems~\cite{Bartolozzi_Indiveri09}. In addition, the
  approach of building distributed multi-chip systems interfaced among
  each other via the \ac{AER} protocol (e.g., see
  Section~\ref{sec:distr-multi-chip}), lends itself well to the
  adoption of event-based mismatch reduction techniques, such as the
  one proposed in~\cite{Neftci_Indiveri10}, that can be effective even
  for very large-scale systems, (e.g., comprising 1 million silicon
  neurons)~\cite{Choudhary_etal12}. In addition to being useful for
  compensating mismatch effects across neurons, homeostatic synaptic
  scaling circuits, such as the ones described in
  Section~\ref{sec:dpiscaling}, can provide another approach to
  compensating the effects of temperature drifts, complementing
  dedicated sub-threshold bias generator
  approaches~\cite{Delbruck_Van-Schaik05,Delbruck_etal10}.  In
  summary, this neuromorphic approach makes it possible to tolerate
  noise, temperature, and mismatch effects at the single device level
  by exploiting the adaptive features of the circuits and
  architectures designed,  leading to robustness at the system level.}{}

\subsection{Exploiting variability and imprecision}
\label{sec:expl-vari-impr}
The strategy proposed by this approach essentially advocates the
construction of distributed and massively parallel computing systems
by integrating very compact, but inaccurate and in\-homogeneous
circuits into large dense arrays, rather than designing systems based
on small numbers of very precise, but large and homogeneous computing
elements. Indeed, intrinsic variability and diverse activation
patterns are often identified as fundamental aspects of neural
computation for information maximization and
transmission~\cite{Maass_etal02, Rigotti_etal10, Shew_etal11,
  Schneidman_etal03}. The strategy of combining large numbers of
variable and imprecise  computing elements to carry out
robust computation is also \revised{variability}{followed
  by}{compatible with} a wide set of traditional machine learning
approaches. These approaches work on the principle of combining the
output of multiple inaccurate computational modules that have slightly
different properties, to optimize classification performances and
achieve or even beat the performances of single accurate and complex
learning systems~\cite{Jacobs_etal91, Breiman01}.  A set of similar
theoretical studies showed that the coexistence of multiple
\revised{timescales}{different}{} time-scales of synaptic plasticity
(e.g., present due to mismatch in the time-constants of the \ac{DPI}
synapse circuits) can dramatically improve the memory performance of
\ac{ANN}~\cite{Fusi_etal05}. The coexistence of slow and fast learning
processes has been shown to be crucial for reproducing the flexible
behavior of animals in context-dependent decision-making (i.e.,
cognitive) tasks and the corresponding single cell recordings in a
neural network model~\cite{Fusi_etal07}.

\subsection{Towards autonomous cognitive systems}
\label{sec:towards-auton-cogn}
Building cognitive systems using noisy and in\-homo\-geneous
subthreshold analog \ac{VLSI} circuits might appear as a daunting
task. The neural circuits and architectures presented in this paper
represent a useful set of building blocks paving the way toward this
goal.  These circuits, as well as analogous one proposed in the
literature~\cite{Indiveri_etal11}, have been used to build compact,
low-power, scalable, computing systems that can interact with the
environment~\cite{Silver_etal07,Neftci_etal10,Choudhary_etal12}, learn
about the input signals they have been designed to
process~\cite{Mitra_etal09}, and exhibit adaptive abilities analogous
to those of the biological systems they
model~\cite{Indiveri03b,Bartolozzi_Indiveri09,Mill_etal11}.  We showed
in this paper how the \ac{sWTA} networks and circuits presented can
implement models of working memory and decision making, thanks to
their selective amplification and reverberating activity properties,
which are often associated to high-level cognitive
abilities~\cite{Eliasmith_etal12}.  Multi-chip systems employing these
architectures can reproduce the results of a diverse set of
theoretical studies based on models of \ac{sWTA} and \ac{ANN} to
demonstrate cognitive properties: for example, Sch\"oner and
Sandamirskaya~\cite{Schoner08,Sandamirskaya_Schoner10} link the types
of neural dynamics described in Section~\ref{sec:networks} to
cognition by applying similar network architectures to sensory-motor
processes and sequence generation; Rutishauser and
Douglas~\cite{Rutishauser_Douglas09} show how the \ac{sWTA} networks
described in this paper can be configured to implement finite state
machines and conditional branching between behavioral
states~\cite{Neftci_etal12b}; Rigotti and
colleagues~\cite{Rigotti_etal10, Rigotti_etal10b} describe neural
principles, compatible with the ones implemented by the circuits
described in Section~\ref{sec:lcircuits}, for constructing recurrent
neural networks able to produce context-dependent behavioral
responses; Giulioni and colleagues~\cite{Giulioni_etal12} demonstrate
working memory in a spiking neural network implemented using the same
type of silicon neuron circuits and plasticity
mechanisms~\cite{Giulioni_etal08} described in
Sections~\ref{sec:silicon-neurons} and~\ref{sec:lcircuits}.

\revised{eliasmith2}{We recently demonstrated how the circuits and
  networks presented in
  Sections~\ref{sec:silicon-neurons},~\ref{sec:silicon-synapses},
  and~\ref{sec:networks} can be used to synthesize cognition on neural
  processing systems~\cite{Neftci_etal13}. Specifically, the
  neuromorphic multi-chip system proposed was used to carry out a
  context-dependent task selection procedure, analogous to the
  sensory-motor tasks adopted to probe cognition in primates.  This is
  a concrete example showing how neuromorphic systems, built using
  variable and imprecise circuits, can indeed be configured to express
  cognitive abilities comparable to those described
  in~\cite{Eliasmith_etal12,Rigotti_etal10}.}{}

\subsection{Challenges and progress in Neuromorphic Engineering}
\label{sec:chall-progr-neur}
Many years have passed since the first publication on neuromorphic
electronic systems~\cite{Mead90}, and remarkable progress has been
made by the small but vibrant \ac{NME}
community~\cite{Telluride,CapoCaccia}. For example the \ac{NME}
community has mastered the art of building real-time sensory-motor
reactive systems, by interfacing circuits and networks of the type
described in this paper with neuromorphic event-based
sensors~\cite{Liu_Delbruck10}; new promising neural-based approaches
have been proposed that link neuromorphic systems to machine
learning~\cite{Nessler_etal09,Steimer_etal09,Corneil_etal12,OConnor_etal13,Neftci_etal14};
substantial progress has been made in the field of
\revised{robots}{neuromorphic}{autonomous}
robots~\cite{Krichmar_Wagatsuma11}; and we are now able to engineer
\revised{scaling1}{both large scale neuromorphic systems (e.g., that
  comprise of the order of $10^6$ neurons~\cite{Merolla_etal14})}{}
and complex multi-chip neuromorphic systems (e.g., that can exhibit
cognitive abilities~\cite{Neftci_etal13}). However, compared to the
progress made in more conventional standard engineering and technology
fields, the rate of progress in \ac{NME} might appear to be
disappointingly small. On one hand, this is due to the fact that
\ac{NME} is still a small community involving a small number of
research groups worldwide (e.g., compared to the number of engineers
that are assigned to the industrial development of new \acp{GPU} or
\acp{CPU}), which lacks the technological infrastructure for
automatized design, verification and configuration tools available for
conventional digital \ac{IC} development. On the other hand,
\revised{scaling2}{scaling and engineering challenges are not the main
  issue:}{} the major limiting factor that hinders the fast
development of neuromorphic engineering is related to our limited
understanding of brain function and neural computation, a concept that
Carver Mead himself highlighted already over 20 years ago in a video
interview (that we transcribe here):
\begin{quote}
  ``I think at the present time we have enough technology to build
  anything we could imagine. Our problem is, we do not know what to
  imagine. We don't understand enough about how the nervous system
  computes to really make more complete thinking systems.''
\end{quote}
Progress on theoretical and computational neuroscience is accelerating
dramatically, also thanks to large-scale funding initiatives recently
announced in both Europe and the United
States~\cite{Alivisatos_etal12, Markram12}. At the same time, an
increasing number of companies is beginning to support research and
development in brain-inspired computing
technologies~\cite{McQuinn_etal13, IBM-Cognitive-Computing,
  Samsung-GRO, Brain-Corp-Technology}. Supported by these new
initiatives, progress in \ac{NME} is beginning to accelerate as
well~\cite{Indiveri_Horiuchi11}. In this perspective, reaching the
ambitious goal of building autonomous neuromorphic systems able to
interact with the environment in real-time and to express cognitive
abilities is within the realm of possibility. To \revised{goal}{reach
  this goal}{this end}, however, it is important to follow a truly
multi-disciplinary approach where neuromorphic engineering serves as a
medium for the exploration of robust principles of brain computation
and not only as a technology platform for the simulation of
\revised{simulation}{neuroscience}{existing} models.

\section{Conclusions}
In this paper we proposed circuit and system solutions following the
neuromorphic approach originally proposed in~\cite{Mead90} for
building autonomous neuromorphic cognitive systems. We presented an
in-depth review of such types of circuits and systems, with tutorial
demonstrations of how to model neural dynamics in analog \ac{VLSI}. We
discussed the problems that arise when attempting to implement
spike-based learning mechanisms in physical systems and proposed
circuit solutions for solving such problems. We described examples of
recurrent neural network implementations that can be used to implement
decision making and working-memory mechanisms, and argued how,
together with the circuits described in the previous sections, they
can be used to implement cognitive architectures. We discussed about
the advantages and disadvantages of the approach followed (e.g., for
the subthreshold regime of operation or for mismatch in analog
subthreshold circuits), and proposed system-level solutions that are
inspired by the strategies used in biological nervous
systems. Finally, we provided an assessment of the progress made in
the \ac{NME} field so far and proposed strategies for accelerating it
and reaching the ambitious goal of building autonomous neuromorphic
cognitive systems.

\section*{Acknowledgments}
Many of the circuits and concepts presented here were inspired by the
ideas and work of Rodney Douglas, Misha Mahowald, Kevan Martin,
Matthew Cook, and Stefano Fusi. The \ac{HW}/\ac{SW} infrastructure
used to characterize the chips throughout the years and build
multi-chip systems was developed in collaboration with Paolo Del
Giudice, Vittorio Dante, Adrian Whatley, Emre Neftci, Daniel Fasnacht,
and Sadique Sheik. We acknowledge also Tobi Delbruck, Shih-Chii Liu
and all our other colleagues at the Institute of Neuroinformatics for
fruitful discussions and collaborations.
We would like to thank the reviewers for their constructive comments.
This work was supported by
the EU ERC Grant ``neuroP'' (257219), the EU FET Grant ``SI-CODE'' (284553),
and by the Excellence Cluster 227 (CITEC, Bielefeld University).

\bibliographystyle{IEEEtran}

\begin{IEEEbiography}[{\includegraphics[width=1in,height=1.25in,clip,keepaspectratio]{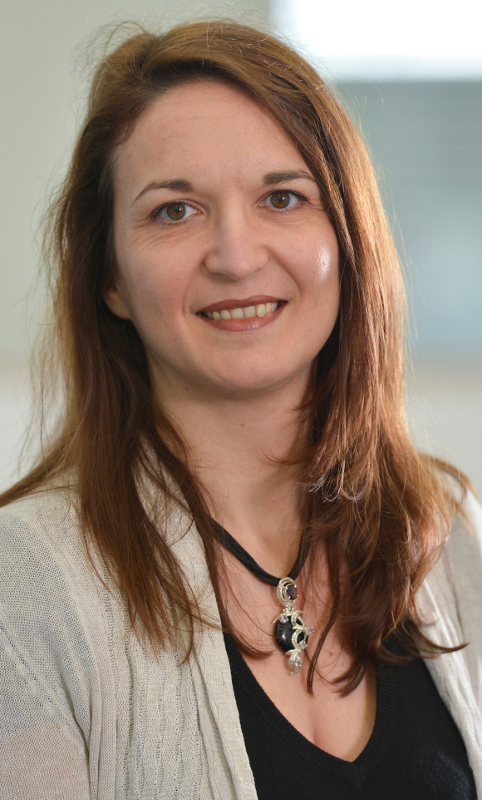}}]{Elisabetta Chicca}
(Member, IEEE) studied physics at the University of Rome 1 ‘La Sapienza’‚ Italy, where she graduated in 1999. In 2006 she received a PhD in Natural Sciences from the Physics department of the Federal Institute of Technology Zurich (ETHZ), Switzerland, and a PhD in Neuroscience from the Neuroscience Center Zurich (ZNZ). Immediately after the PhD, she started a PostDoc at the Institute of Neuroinformatics at the University of Zurich and ETH Zurich, where she continued working as Research Group Leader from May 2010 to August 2011. Since August 2011, she is an assistant professor at Bielefeld University and is heading the Neuromorphic Behaving Systems Group affiliated to the Faculty of Technology and the Cognitive Interaction Technology - Center of Excellence (CITEC). Her current interests
are in the development of VLSI models of cortical circuits for brain-inspired computation, learning in
spiking VLSI neural networks, bio-inspired sensing (olfaction, active electrolocation, audition).

Elisabetta Chicca is member of the IEEE Biomedical Circuits and Systems TC and IEEE Neural
Systems and Applications TC (currently Secretary).
\end{IEEEbiography}

\begin{IEEEbiography}[{\includegraphics[width=1in,height=1.25in,clip,keepaspectratio]{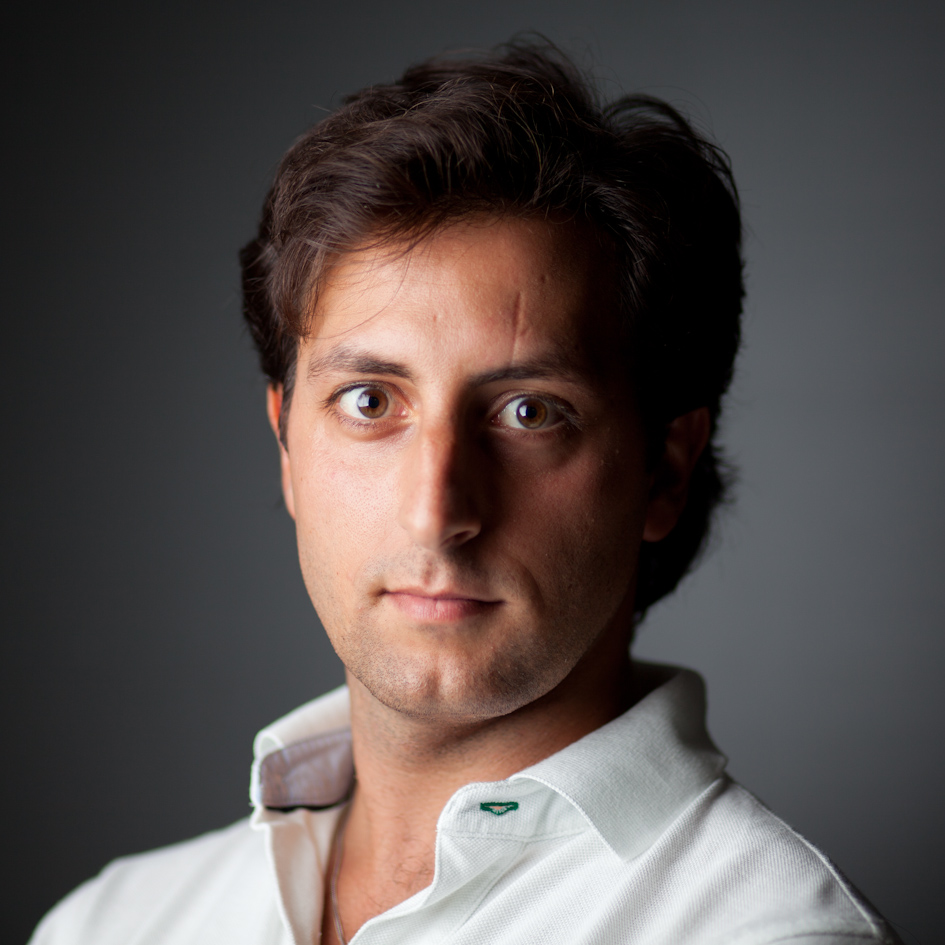}}]{Fabio Stefanini} obtained a Laurea Triennale degree (B.S.) and a "Laurea Magistrale" degree (M.S.) in Physics from La Sapienza University of Rome (Italy) in 2006 and 2009 respectively.
He has been a Research Collaborator at the Institute for Complex Systems, CNR-INFM (Rome, Italy), developing experimental, software and theoretical methods for the study of collective behaviour in flocking birds.
He obtained a Ph.D. at the Institute of Neuroinformatics in Zurich (Switzerland) implementing a brain-inspired, real-time pattern recognition system using neuromorphic hardware with distributed synaptic plasticity.
His main research interests are in neuromorphic systems with analog VLSI circuits, learning neural networks and complex systems.
He currently owns a PostDoc position at the Institute of Neuroinformatics of Zurich.
His research involves the development of cortical-inspired smart processing systems for context-aware, embedded processors for resource management in mobile devices.
Dr. Fabio Stefanini is one of the creators of PyNCS, a Python package proposed as a flexible, kernel-like infrastructure for neuromorphic systems.

\end{IEEEbiography}

\begin{IEEEbiography}[{\includegraphics[width=1in,height=1.25in,clip,keepaspectratio]{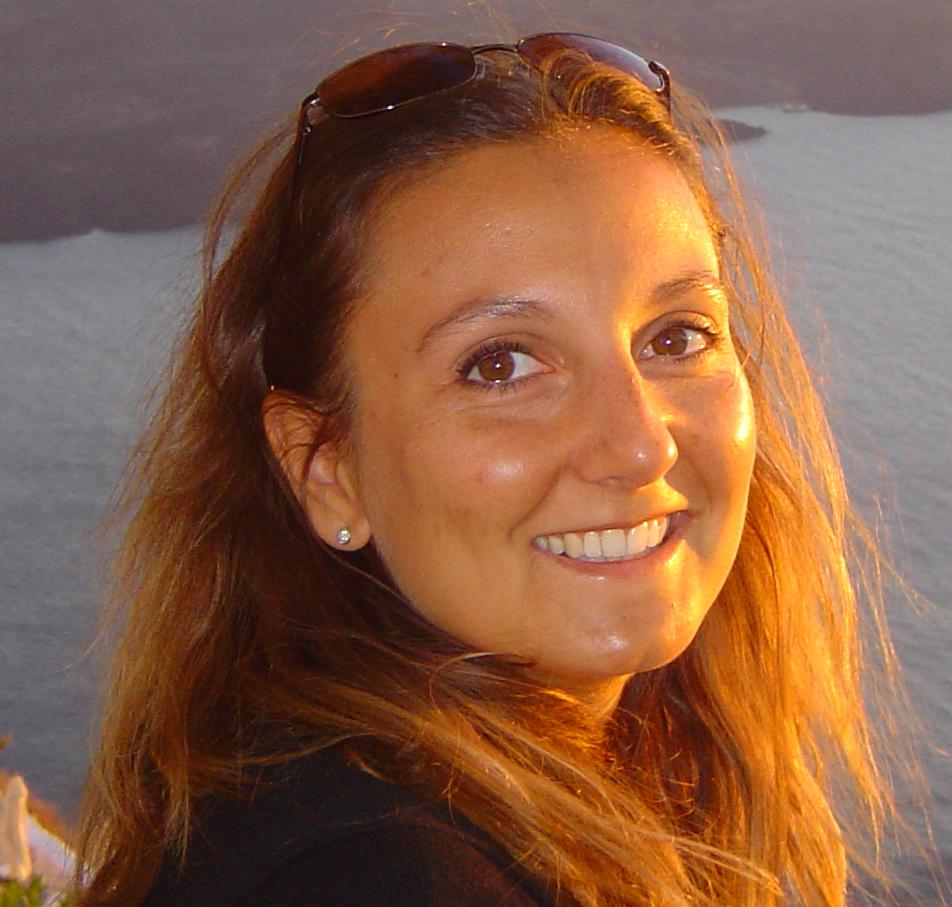}}]{Chiara Bartolozzi}
(Member, IEEE) received the Laurea (with honors) degree in biomedical engineering from the University of Genova, Genova, Italy, in 2001 and the Ph.D. degree in Natural Sciences from the Physics department of the Federal Institute of Technology Zurich (ETHZ), Switzerland, and a PhD in Neuroscience from the Neuroscience Center Zurich (ZNZ) in 2007. She then joined the the Istituto Italiano di Tecnologia, Genova, Italy, first as a PostDoc in the Robotics, Brain and Cognitive Sciences Department and then as Researcher in the iCub Facility, where she is heading the Neuromorphic Systems and Interfaces group. Her main research interest is the design of event-driven technology and their exploitation for the development of novel robotic platforms. To this aim, she coordinated the eMorph (ICT-FET 231467) project that delivered the unique neuromorphic iCub humanoid platform, developing both the hardware integration and the computational framework for event-driven robotics.
She is member of the IEEE Circuits and Systems Society (CASS) Sensory Systems (SSTC) and Neural Systems and Applications (NSA) Committees.
\end{IEEEbiography}

\begin{IEEEbiography}[{\includegraphics[width=1in,height=1.25in,clip,keepaspectratio]{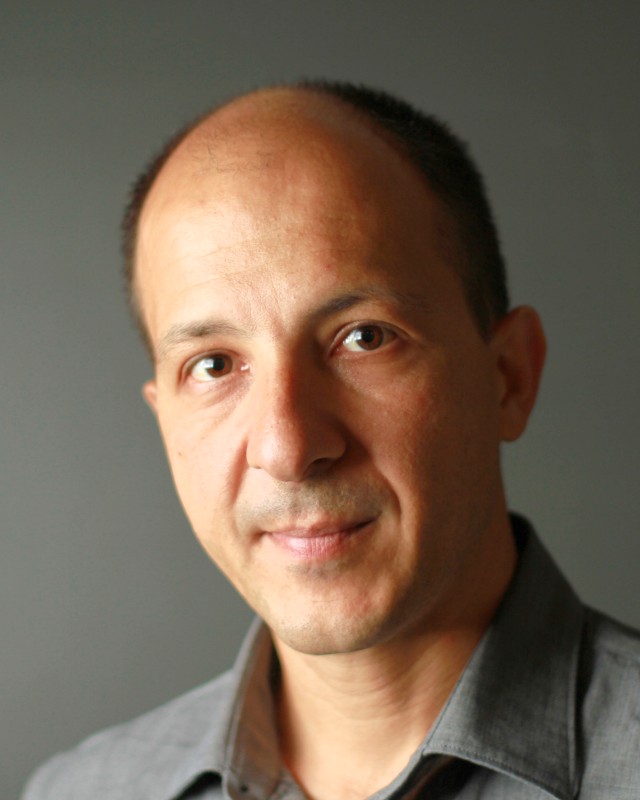}}]{Giacomo Indiveri}
(Senior Member, IEEE) is an Associate Professor at the Faculty of Science, University of Zurich, Switzerland. Indiveri received the M.Sc. degree in electrical engineering from the University of Genoa, Italy in 1992. Subsequently, he was awarded a doctoral postgraduate fellowship within the National Research and Training Program on ``Technologies for Bioelectronics'' from which he graduated with ``summa cum laude'' in 1995. He also obtained a Ph.D. degree in computer science and electrical engineering from the University of Genoa, Italy in 2004, and the ``Habilitation'' certificate in Neuromorphic Engineering from ETH Zurich, Switzerland in 2006.
Indiveri carried out research on neuromorphic vision sensors as a Postdoctoral Research Fellow in the Division of Biology at the California Institute of Technology, Pasadena, CA, USA, and on neuromorphic selective attention systems as a postdoc at the Institute of Neuroinformatics of the University of Zurich and ETH Zurich, Switzerland.  His current research interests lie in the study of real and artificial neural processing systems, and in the hardware implementation of neuromorphic cognitive systems, using full custom analog and digital VLSI technology. Indiveri is a member of several Technical Committees (TCs) of the IEEE Circuits and Systems society and a Fellow of the European Research Council. 
\end{IEEEbiography}

\end{document}